# On the second law of thermodynamics: a global flow spontaneously induced by a locally nonchaotic energy barrier

Yu Qiao,[1,2,*] Zhaoru Shang[1]
[1] *Program of Materials Science and Engineering, University of California – San Diego, La Jolla, CA 92093, U.S.A.*
[2] *Department of Structural Engineering, University of California – San Diego, La Jolla, CA 92093-0085, U.S.A.*
[*] Email: *yqiao@ucsd.edu*

**Abstract:** People are well aware that, inherently, certain small-scale nonchaotic particle movements are not governed by thermodynamics. Usually, such phenomena are studied by kinetic theory and their energy properties are considered "trivial". In current research, we show that, beyond the boundary of the second law of thermodynamics where Boltzmann's H-theorem does not apply, there are also large-sized systems of nontrivial energy properties: when the system is isolated, entropy can decrease; from a single thermal reservoir, the system can absorb heat and produce useful work without any other effect. The key concept is local nonchaoticity, demonstrated by using a narrow energy barrier. The barrier width is much less than the nominal particle mean free path, so that inside the barrier, the particle-particle collisions are sparse and the particle trajectories tend to be locally nonchaotic. Across the barrier, the steady-state particle flux ratio is intrinsically in a non-Boltzmann form. With a step-ramp structure, the nonequilibrium effect spreads to the entire system, and a global flow is generated spontaneously from the random thermal motion. The deviation from thermodynamic equilibrium is steady and significant, and compatible with the basic principle of maximum entropy. These theoretical and numerical analyses may shed light on the fundamentals of thermodynamics and statistical mechanics.

## 1. Introduction

The second law of thermodynamics is critical to many fields in physics, including quantum mechanics, astrophysics, materials science, energy science, etc. [1]. Yet, unlike the first law of thermodynamics that is entailed by Noether's theorem, it does not have a solid theoretical underpinning [2]. In the classic H-theorem [3], Boltzmann intended to answer the following question: in an isolated system, as all the governing equations are time-reversible, how can entropy increase be irreversible? He brought in the assumption of molecular chaos, which requires extensive particle-particle collisions throughout the entire system. Currently, no decisive conclusion has been reached if in a large-sized system, while ergodic and chaotic in most areas, the global state is dominated by a local process of negligible particle interaction.

Over the years, there were continued efforts to study the "counterexamples" of the second law of thermodynamics. They have hitherto demonstrated the robustness of the theory of statistical mechanics, which inspired the investigation on nonequilibrium and stochastic thermodynamics [e.g., 4]. In general, these works can be represented by two classical models: Maxwell's demon [e.g., 5,6] and Feynman's ratchet [7]. Both of them have a variety of variants. For example, Maxwell's demon can operate the Szilárd engine [8]; Feynman's ratchet is somewhat equivalent to Smoluchowski's trapdoor [9] and the "autonomous Maxwell's demon" (the single-electron refrigerator) [10]. Maxwell's demon is nonequilibrium, but not spontaneous; Feynman's ratchet is spontaneous, but not nonequilibrium. Maxwell's demon relies on external intervention and therefore, the entropy reduction is associated with additional information [11,12] and an energetic penalty [13,14]; in Feynman's ratchet, the time-average behaviors of all the components are balanced [7].

Recently, we investigated the concept of spontaneously nonequilibrium dimension (SND), and concluded that an SND-based system may not be governed by the second law of thermodynamics [15,16]. In general, if across an area, inherently, the particle number density distribution cannot reach thermodynamic equilibrium, we refer to this area as a SND. It combines the nonequilibrium characteristics of Maxwell's demon with the spontaneity of Feynman's ratchet.

In an isolated large-sized system, the second law of thermodynamics forbids the existence of SND. Yet, our study suggested otherwise. One example of SND is a narrow energy barrier with a width much less than the particle mean free path, so that the particle-particle collisions are sparse inside the barrier [15]. The numerical analysis indicated that useful work could be produced in a cycle by absorbing heat from a single thermal reservoir, which was attributed to the asymmetry in the cross-influence of thermally correlated thermodynamic forces. Motivated by this finding, we designed and carried out an experiment on an entropy-barrier SND [16]. The testing data demonstrated entropy decrease without an energetic penalty. To adapt to these remarkable phenomena and also remain consistent with the basic principle of maximum entropy, the second law of thermodynamics can be generalized as $S \rightarrow S_{\mathrm{Q}}$ [16]: in an isolated system, the difference between entropy ($S$) and the maximum possible steady-state value ($S_{\mathrm{Q}}$) cannot increase. If $S_{\mathrm{Q}}$ is equal to the equilibrium maximum ($S_{\mathrm{eq}}$), $S \rightarrow S_{\mathrm{Q}}$ converges to the entropy statement of the second law of thermodynamics, i.e., entropy of an isolated system can never decrease. When $S_{\mathrm{Q}}$ is reduced by the SND to the nonequilibrium maximum ($S_{\mathrm{ne}}$), $S$ may decrease.

The operation of the previous model in [15] is relatively complicated, and the analysis is taxing and contains errors (see Section 4.4). The parameters must be alternately adjusted. Below, we design and investigate another model system of SND. The system configuration is quite simple, and the primary procedure is autonomous. Section 2 presents a preparatory study on a Knudsen gas in gravity (a narrow energy barrier), showing how nonchaoticity can spontaneously render the steady state non-Boltzmannian. By itself, this phenomenon does not directly conflict with any thermodynamic laws, as the Knudsen gas is small in height and is not a thermodynamic system. Sections 3 and 4 present the large-sized ideal-gas model with a step-ramp structure. The low-height step is similar to the Knudsen-gas setup in Section 2. Section 5 presents a reanalysis of the numerical findings in [15], with a few new components; it may be helpful for further understanding the step-ramp model in Sections 3-4.

In this manuscript, the term "nonchaotic" (or "locally nonchaotic") is used to describe an area wherein particle-particle collisions are sparse, so that the particle trajectories inside the area tend to be nonchaotic, i.e., a small change in the initial condition would not cause a drastically different evolution path. The term "nonequilibrium" (or "intrinsically nonequilibrium") describes a steady state significantly different from thermodynamic equilibrium; the system is large-sized, either isolated from exchanging energy and mass with the environment, or closed and immersed in a thermal bath. Unless otherwise specified, we are not interested in fluctuation or transient system behaviors. The analysis is in the framework of classical mechanics.

## 2. Preparatory Analysis: Nonchaotic Particle Movement in a Narrow Energy Barrier

In this section, we perform a kinetic analysis on a Knudsen gas in a uniform gravitational field, preparatory to the thermodynamic analysis of the ideal-gas model. We examine the nonchaotic particle movement in the narrow energy barrier, and show that the steady-state particle flux ratio ($\delta$) is inherently non-Boltzmannian. A similar energy barrier will be employed as the key component in the large-sized step-ramp system in Section 3.

### 2.1 Low-height vertical plane with billiard-like particles

Figure 1(a) depicts a vertical y-z plane, wherein a large number of elastic particles randomly move. A uniform gravitational field ($g$) is in the -z direction. In this two-dimensional (2D) system, the particles are finite-sized hard disks. There is no long-range force among them.

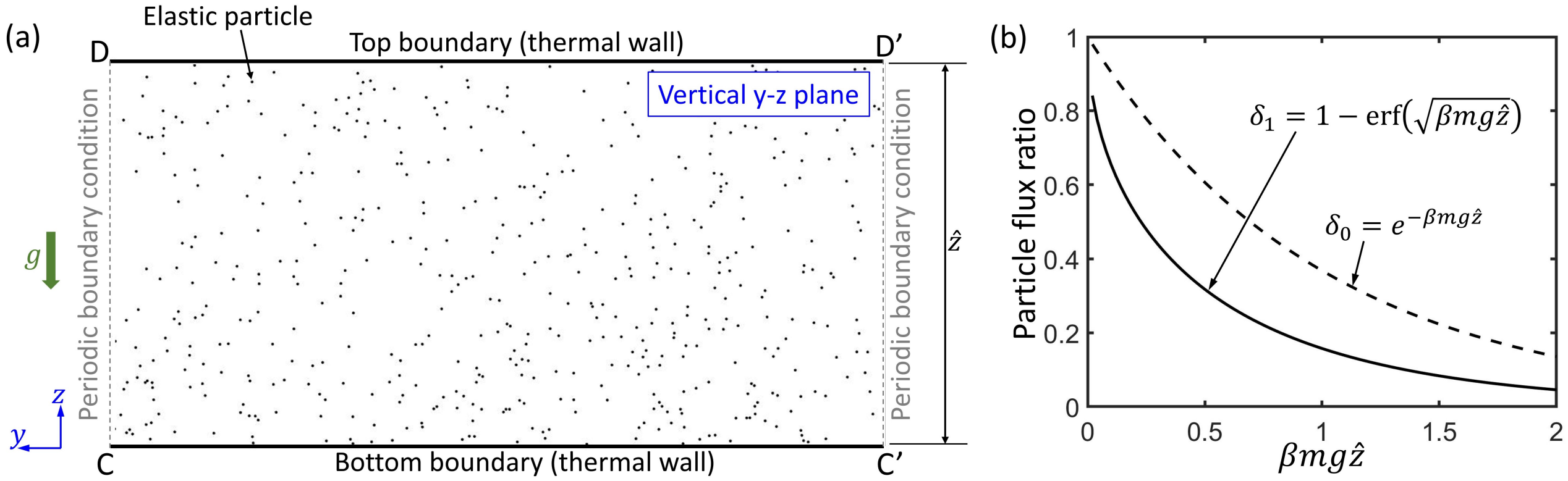

**Fig. 1 (a)** A vertical y-z plane, in which a large number of two-dimensional elastic particles randomly move in a gravitational field ($g$). When the plane height ($\hat{z}$) is less than the nominal particle mean free path ($\lambda_{\mathrm{F}}$), the system can be viewed as a Knudsen gas. The left and right borders are open and use periodic boundary condition; the upper and lower borders are thermal walls at the same temperature. **(b)** Comparison between the Boltzmann factor ($\delta_0$) and the non-Boltzmann steady-state particle flux ratio ($\delta_1$), as functions of the normalized energy barrier ($\beta m g \hat{z}$).

The lateral borders (DC and D′C′) are open and use periodic boundary condition. The top and bottom boundaries (DD′ and CC′) are thermal walls at the same temperature, from which the reflected particle direction is random; the speed of the reflected particles is not correlated with the incident speed, but randomly follows the 2D Maxwell-Boltzmann distribution, $p(v) = (mv/\overline{K})e^{-mv^2/(2\overline{K})}$, with $v$ being the particle speed ($v > 0$), $m$ the particle mass, $\overline{K} = k_{\mathrm{B}}T$, $k_{\mathrm{B}}$ the Boltzmann constant, and $T$ the nominal temperature. The nominal temperature is mainly used as a parameter to set the boundary condition and initial condition in the computer simulation.

The condition of nonchaoticity (i.e., negligible particle-particle collision) can be stated as $\hat{z} \ll \lambda_{\mathrm{F}}$, where $\hat{z}$ is the plane height, $\lambda_{\mathrm{F}} = A_0/\left(\sqrt{8}Nd\right)$ is the nominal particle mean free path, $A_0$ is the area of particle movement, $N$ is the total particle number, and $d$ is the particle diameter. When a particle moves in the vertical plane, the maximum possible time interval between particle-wall collisions is $2\tau_0 = 2\sqrt{2\hat{z}/g}$, and the associated characteristic horizontal particle displacement is $\bar{y}_{\mathrm{m}} = 2\tau_0/\sqrt{\beta m} = 2\sqrt{2/\eta}\,\hat{z}$, where $\beta = 1/(k_{\mathrm{B}}T)$, and $\eta = \beta m g \hat{z}$ is the normalized energy barrier. With a given $\eta$, as long as $\hat{z}$ is sufficiently small, $\bar{y}_{\mathrm{m}} \ll \lambda_{\mathrm{F}}$. Under this condition, the particle-particle collisions in Figure 1(a) are sparse, and the system state is dominated by the particle-wall collisions at DD′ and CC′. As a particle ascends, to overcome the gravitational energy barrier, the horizontal component of particle momentum ($p_{\mathrm{y}}$) has nearly no contribution; only the z-direction kinetic energy ($K_{\mathrm{z}} \triangleq m v_{\mathrm{z}}^2/2$) is important, where $v_{\mathrm{z}}$ is the z-component of particle velocity.

Define the particle flux ratio ($\delta$) as the steady-state $n_{\mathrm{t}}/n_{\mathrm{b}}$ ratio, where $n_{\mathrm{t}}$ and $n_{\mathrm{b}}$ are the numbers of particle-wall collisions at the top boundary DD′ ($z = \hat{z}$) and the bottom boundary CC′ ($z = 0$), respectively. For the nonchaotic setup ($\hat{z} \ll \lambda_{\mathrm{F}}$), $\delta$ may be estimated as*

$$\delta_1 = \int_{\sqrt{2g\hat{z}}}^{\infty} p_{\mathrm{z}}(v_{\mathrm{z}})\,\mathrm{d}v_{\mathrm{z}} = 1 - \mathrm{erf}\left(\sqrt{\beta m g \hat{z}}\right) \tag{1}$$

where $p_{\mathrm{z}}(v_{\mathrm{z}}) = \sqrt{2m/(\pi\overline{K})}\, e^{-m v_{\mathrm{z}}^2/(2\overline{K})}$ is the one-dimensional Maxwell-Boltzmann distribution of $|v_{\mathrm{z}}|$. In general, $\delta_1$ is smaller than the Boltzmann factor, $\delta_0 = e^{-\beta m g \hat{z}}$ (see Figure 1b). Only when $\hat{z}/\lambda_{\mathrm{F}} \gg 1$, with extensive particle-particle collision, could the system reach thermodynamic equilibrium, i.e., $\delta \to \delta_0$. Notice that $\delta$ is different from the particle number density distribution, due to the nonuniformity of $v_{\mathrm{z}}$. Section A1 in the Appendix gives more discussions on how, without extensive particle-particle collision, $\delta$ is non-Boltzmannian.

There are a couple of points worth noting. Firstly, Figure 1(a) is a simplified analog to the vertical step in the ideal-gas model to be analyzed in Section 3. The boundary condition at the

* The crossing ratio ($\delta$) may be better assessed as $\delta_{\mathrm{k}} = \int_{\sqrt{2g\hat{z}}}^{\infty}[\bar{v}_{\mathrm{z}} \cdot p_{\mathrm{z}}(v_{\mathrm{z}})]\mathrm{d}v_{\mathrm{z}} / \int_0^{\infty}[\bar{v}_{\mathrm{D}} \cdot p_{\mathrm{D}}(v_{\mathrm{z}})]\mathrm{d}v_{\mathrm{z}}$, where $\bar{v}_{\mathrm{z}} = \hat{z}/\bar{t}_{\mathrm{v}}$ is the average $v_{\mathrm{z}}$ along the vertical axis, $\bar{t}_{\mathrm{v}} = \left(v_{\mathrm{z}} - \sqrt{v_{\mathrm{z}}^2 - 2g\hat{z}}\right)/g$ is the time it takes for a particle to ascend from the bottom wall to the top wall, $p_{\mathrm{D}}$ is the 1D Maxwell-Boltzmann distribution of $|v_{\mathrm{z}}|$ of the descending particles ($p_{\mathrm{D}} = p_{\mathrm{z}}$ in Figure 1A), $\bar{v}_{\mathrm{D}} = \hat{z}/\bar{t}_{\mathrm{D}}$, and $\bar{t}_{\mathrm{D}} = \left(\sqrt{v_{\mathrm{z}}^2 + 2g\hat{z}} - v_{\mathrm{z}}\right)/g$ is the descending time. There is no analytical solution of $\delta_{\mathrm{k}}$. With the parameter ranges in this manuscript, the difference between $\delta_{\mathrm{k}}$ and $\delta_1$ is on the scale of 10%.

upper and lower thermal walls approximately represents the chaotic gas in the large horizontal areas (the plateau and the plain) in Figure 3(a), as discussed in Section A2 in the Appendix. The two walls are set to the same $T$, which ignores the heterogeneous particle velocity distribution (see Section 4.3 below). Secondly, to obtain an appreciable difference between $\delta_0$ and $\delta_1$, $\eta \triangleq \beta m g \hat{z}$ must be nontrivial. If the particles are air molecules, $m \approx 29$ g/mol and $\lambda_{\mathrm{F}} \approx 68$ nm. At room temperature, to render $\eta$ larger than 0.01, with $\hat{z} \approx 5$ nm, $g$ needs to be on the scale of $10^{11}$ m/s$^2$, at the level of neutron stars. It imposes tough challenges to the direct experimental verification. Nonetheless, the current theoretical study may shed light on the fundamentals of thermodynamics, and inspire future experiments using entropy barriers [16] or stronger thermodynamic forces (see Section 4.4).

## 2.2 Monte Carlo simulation

The influence of $\hat{z}$ on $\delta$ is visualized by a Monte Carlo (MC) simulation (Figure 1a). The computer program is available at [17]; the algorithm is summarized in Section A3 in the Appendix. The setup is scalable; an example of the unit system can be based on K, g/mole, Å, and fs. The particle diameter ($d$) is 1; the area of particle movement is $A_0 = \hat{z} \cdot w_0 = 39268.75$, where $\hat{z} = h - d$, and $h$ and $w_0$ are the height and the width of the simulation box, respectively; the particle number $N = 500$; $m = 1$; $T = 300$. The nominal particle mean free path $\lambda_{\mathrm{F}} = A_0/\left(\sqrt{8}Nd\right) \approx 27.77$, and the percentage of the occupied area of the particles is $N\pi d^2/(4A_0) \approx 1\%$.

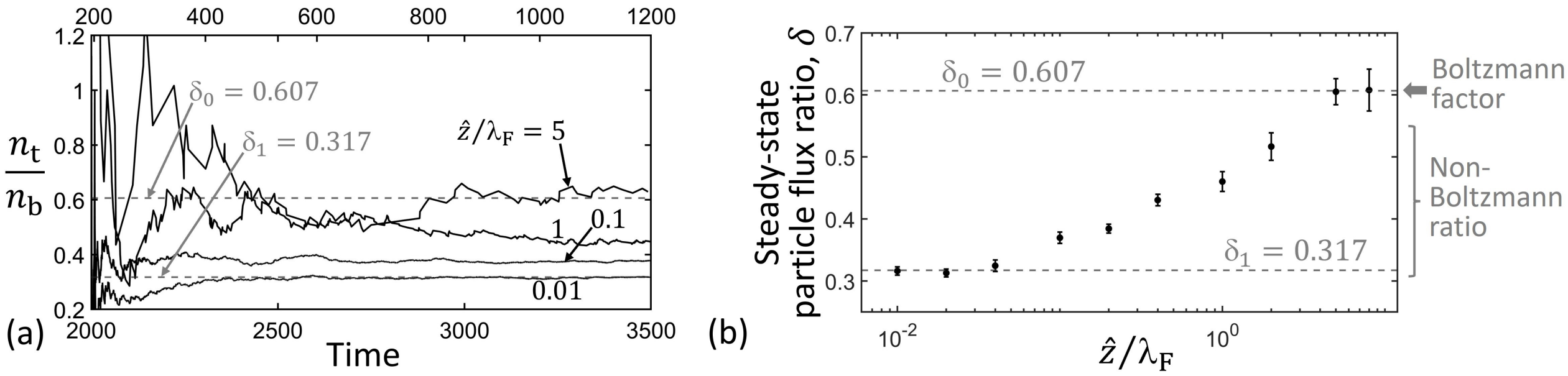

**Figure 2 (a)** Typical time profiles of the $n_{\mathrm{t}}/n_{\mathrm{b}}$ ratio, with $\hat{z}/\lambda_{\mathrm{F}}$ being 5, 1, 0.1, or 0.01, where $n_{\mathrm{t}}$ and $n_{\mathrm{b}}$ are the numbers of the particle-wall collisions at the upper boundary and the lower boundary, respectively; $\hat{z}$ is the plane height, and $\lambda_{\mathrm{F}}$ is the nominal particle mean free path. The upper ruler of the horizontal axis is for $\hat{z}/\lambda_{\mathrm{F}} = 0.01$; the lower ruler is for the other three curves. **(b)** The particle flux ratio ($\delta$) as a function of $\hat{z}/\lambda_{\mathrm{F}}$; $\delta$ is defined as the steady-state $n_{\mathrm{t}}/n_{\mathrm{b}}$ ratio.

In different simulation cases, $\hat{z}$ is varied; $\hat{z}/\lambda_F$ ranges from 0.01 to 8. The width of the simulation box ($w_0$) is changed accordingly, to keep $A_0$ and $\lambda_F$ constant. The gravitational acceleration ($g$) is adjusted to maintain $\beta mg\hat{z} = 0.5$, so that $\delta_0 = e^{-\beta mg\hat{z}} = 0.607$ and $\delta_1 = 1 - \mathrm{erf}\left(\sqrt{\beta mg\hat{z}}\right) = 0.317$ remain unchanged. At time zero, the particles are randomly placed in the simulation box. The probability density function of the initial particle speed is the 2D Maxwell-Boltzmann distribution, $p(v)$. The initial particle direction is random. If $\hat{z}/\lambda_F < 1$, the timestep of simulation ($\Delta t_0$) is set to 0.0183; if $\hat{z}/\lambda_F \geq 1$, $\Delta t_0 = 0.0058$.

For each simulation case, after the settlement period ($t_{sp} = 1.826 \times 10^3$), we begin to count the number of particle-boundary collisions at the upper and lower walls, $n_t$ and $n_b$. Figure 2(a) shows the typical time profiles of the running average of the $n_t/n_b$ ratio. Figure 2(b) shows the steady-state $\delta$ as a function of $\hat{z}/\lambda_F$. For each $\hat{z}/\lambda_F$, three nominally same simulations are carried out, with randomized initial conditions. In this manuscript, the error bars indicate the 90%-confidence interval, $\pm 1.645 \cdot S_t/\sqrt{N_s}$, with $S_t$ being the standard deviation and $N_s$ the number of data points.

In Figure 2(b), when $\hat{z} \gg \lambda_F$ (i.e., the particle-particle interaction is extensive), $\delta \rightarrow \delta_0$, following the classical barometric formula [18]; when $\hat{z} \ll \lambda_F$ (i.e., the particle-particle interaction is negligible), $\delta \rightarrow \delta_1$. Such a $\delta - \hat{z}$ relationship is in agreement with the numerical result in [15]. We tested various settings in the computer simulation (see Section A4 in the Appendix). As long as there is no extensive particle-particle collision ($\hat{z}/\lambda_F \ll 1$), the steady state is always significantly nonequilibrium (i.e., $\delta \neq \delta_0$), regardless of the specific forms of the boundary condition and the initial condition.

When $\hat{z}/\lambda_F < 1$, Figure 1(a) is a Knudsen gas [19]. It has been well known that a Knudsen gas may not reach thermodynamic equilibrium, and it is usually not studied by thermodynamics [20-22]. Compared with conventional Knudsen gases, a unique characteristic of Figure 1(a) and the model system to be studied in Section 3 (Figure 3a) is that the gravitational effect is nontrivial. Moreover, in Figure 3(a), the vertical step is only a small portion of the entire structure, and is fully open on the upper and lower sides. When we perform thermodynamic analysis on Figure 3(a), the large-sized system may be more conveniently treated as an ideal gas, contained in a special-shaped step-ramp container. The system entropy can be calculated by using the entropy equation of ideal gas, as will be discussed in Section 4.1.

## 3. Large-Sized Ideal-Gas Model with a Locally Nonchaotic Energy Barrier

In this section, we investigate a large-sized ideal-gas model, which contains a locally nonchaotic energy-barrier SND similar to the vertical plane in Figure 1(a). It demonstrates the counterintuitive effects of the intrinsically nonequilibrium steady state.

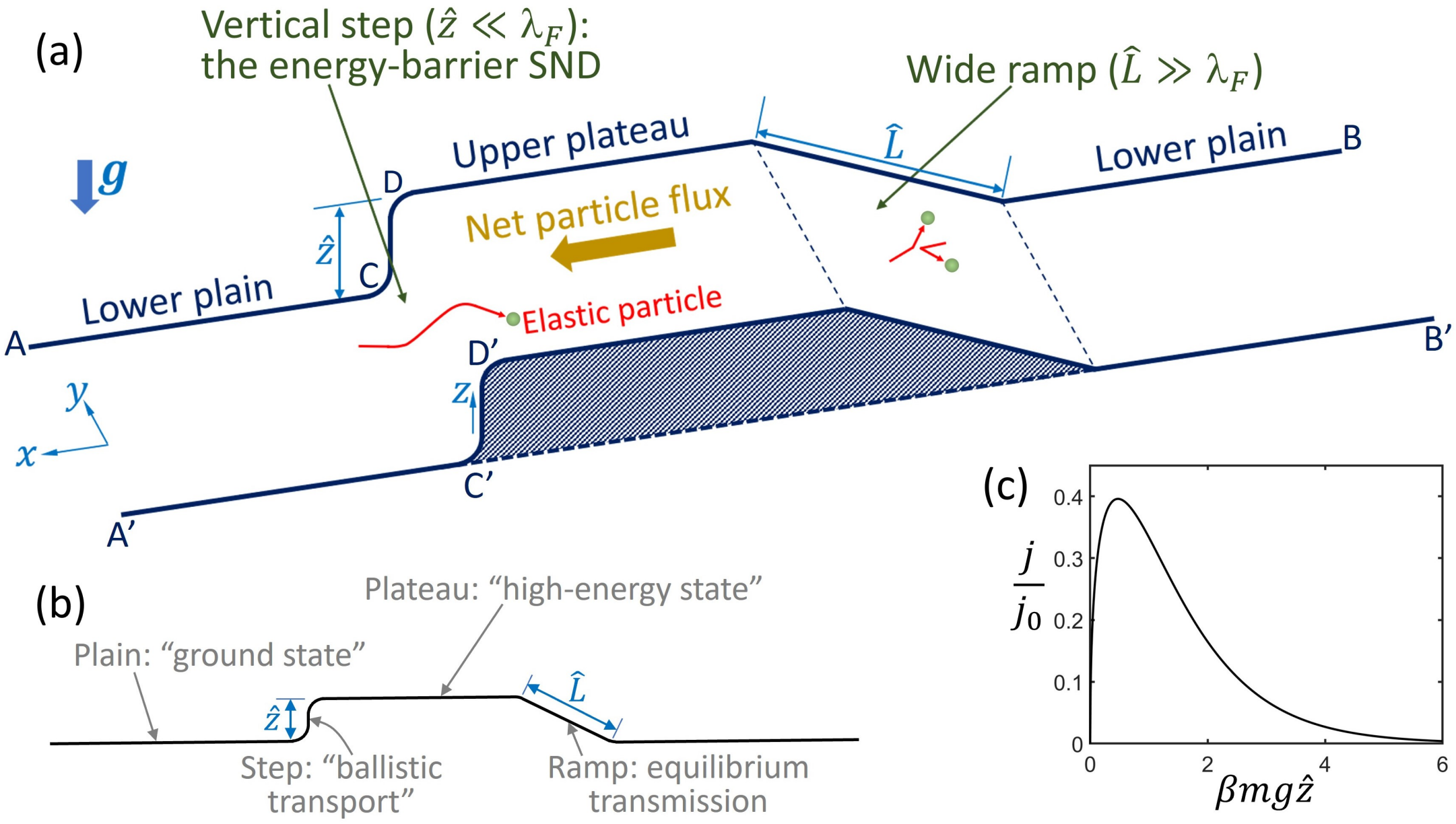

**Fig. 3 (a)** Three-dimensional view and **(b)** side view of the large-sized model system. It can be viewed as an ideal gas in a special-shaped step-ramp container. The front and back borders (AA′ and BB′) are open and use periodic boundary condition. The lateral borders (AB and A′B′) are isolated from the environment. **(c)** An example of the steady-state particle flow rate, $j$, predicted by Equation (2). When the normalized energy barrier ($\beta m g \hat{z}$) is in the middle range, $j$ is nontrivial.

### 3.1 Basic concept: two-ended structure

Figure 3(a,b) depicts the ideal-gas model system. It has a two-ended structure, partly inspired by Feynman’s ratchet [7] (see Section A5 in the Appendix). The central area (the upper “plateau”) is higher than the rest of the area (the lower “plain”). The plain and the plateau can be arbitrarily large, wherein the particle movement is ergodic and chaotic (see Section A2 in the Appendix). The plateau height ($\hat{z}$) is much less than the nominal particle mean free path ($\lambda_{\mathrm{F}}$). A uniform gravitational field ($g$) is along −z, normal to the plain and the plateau. The left-hand side

and the right-hand side of the plateau are connected to the plain through a vertical step and a wide ramp, respectively. The ramp size ($\hat{L}$) is much larger than $\lambda_F$. The front and back borders (AA′ and BB′) are open and use periodic boundary condition. The lateral borders (AB and A′B′) are isolated from the environment.

On the one hand, since $\hat{L} \gg \lambda_F$, the particle collisions in the ramp are extensive. Across the ramp, the particle number density ratio between the plateau and the plain ($\hat{\rho}$) tends to be the Boltzmann factor [3], $\delta_0 = e^{-\beta mg\hat{z}}$. On the other hand, as demonstrated in Figure 2, the locally nonchaotic vertical step may exhibit a non-Boltzmann particle flux ratio ($\delta_*$). As a result, the local $\hat{\rho}$ tends to be smaller than $\delta_0$.

As $\delta_* < \delta_0$, the low-height step behaves as a spontaneously nonequilibrium dimension (SND), and the system could be unbalanced. At the steady state, the overall probability for the particles to move across the ramp along +x is larger than the probability for the particles to move across the step along -x. Therefore, there would be a net particle flux ($j$) in the +x direction (from the ramp side to the step side on the plateau). For the sake of simplicity, here we analyze a system in which the plateau and the plain are much larger than the ramp and the step. The steady-state particle number densities on the plateau ($\rho_G \triangleq N_G/A_G$) and the plain ($\rho_P \triangleq N_P/A_P$) may be assessed through $\rho_G A_G + \rho_P A_P \approx N$ and $\rho_G/\rho_P \approx \bar{\delta}$, where $\bar{\delta} = (\delta_0 + \delta_*)/2$, $N_G$ and $N_P$ are respectively the particle numbers on the plateau and the plain, and $A_G$ and $A_P$ are respectively the areas of the plateau and the plain. Thus, $\rho_P = N/\left(\bar{\delta} A_G + A_P\right)$. As a first-order approximation, the steady-state flow rate can be estimated as

$$j = \frac{1}{2}(\rho_P \delta_0)\bar{v}_x - \frac{1}{2}(\rho_P \delta_*)\bar{v}_x = \frac{1}{2}\frac{\tilde{\rho}}{\bar{\delta}\tilde{A}+1}\Delta\delta \cdot \bar{v}_x \tag{2}$$

where $\tilde{\rho} = N/A_P$, $\Delta\delta = \delta_0 - \delta_*$, $\tilde{A} = A_G/A_P$, and $\bar{v}_x = \sqrt{2k_B T/(\pi m)}$. Accordingly, the steady-state drift velocity on the plain is

$$v_w = \frac{j}{\rho_P} = \frac{1}{2}\Delta\delta \cdot \bar{v}_x \tag{3}$$

Figure 3(c) shows one example of Equation (2), where $\delta_*$ is taken as $\delta_1$, $j_0 = \bar{\rho} \cdot \bar{v}_x/2$, $\bar{\rho} = N/A_0$, $A_0 = A_G + A_P$, and $\tilde{A}$ is set to 1. When $\hat{z} = 0$, the energy barrier vanishes, so that $j = 0$. When the energy barrier is large, because both $\delta_0$ and $\delta_*$ are small, few particles are on the plateau and consequently, $j$ is also near zero. When $\beta mg\hat{z}$ is in the middle range, $j$ is significant.

In essence, the lower plain may be viewed as the "ground state"; the upper plateau represents a "high-energy state". The wide ramp is a path of equilibrium particle transmission. The SND (the low-height vertical step) provides a mechanism of nonequilibrium particle transmission. It plays a somewhat similar role to Maxwell's demon [5] by selectively allowing the high-$v_z$ particles to pass in the -x direction, so that $\delta_* \neq \delta_0$. Yet, the vertical step does not involve any external information processing or active control. The flow comes spontaneously from the unforced thermal movement of particles, not subject to an energetic penalty.

3.2 Monte Carlo simulation: spontaneous particle flow

To demonstrate the concept of Figure 3, we perform a MC simulation on a 2D system. The computer program is available at [17]. The simulation box represents the surface of particle movement (Figure 4a). From left to right, it contains a left plain ("+"), a step, a plateau, a wide ramp, and a right plain ("–"). The simulation is scalable; an example of the unit system can be based on nm, fs, g/mol, and K. The width of the simulation box between AB and A′B′ ($w_0$) is 50. The length of each plain ("+" or "–") is $L_{\mathrm{P}} = 5$. The plateau length ($L_{\mathrm{G}}$) is 10. The step size ($\hat{z}$) is 0.5. The ramp size ($\hat{L}$) is 50. The total particle number $N$ = 500; $d = 0.2$; $m = 1$; the timestep $\Delta t_0 = 1$; $T$ is set to 1000, which is mainly used to compute $\beta$ and $p(v)$ for the initial condition. The nominal particle mean free path is $\lambda_{\mathrm{F}} = A_0/\left(\sqrt{8}Nd\right) \approx 12.46$. Across the step, the local $\lambda_{\mathrm{F}}$ on the plateau tends to be larger than on the plain by a factor of $1/\delta_*$. If $\delta_* \approx \delta_1$, when $\beta m g \hat{z} = 0.2$, $1/\delta_* \approx 1.9$. The minimum local $\lambda_{\mathrm{F}}$ is ~9 (on the plain), much larger than $\hat{z}$; the maximum local $\lambda_{\mathrm{F}}$ is ~17 (on the plateau), considerably smaller than $\hat{L}$.

In the step surface, the gravitational acceleration ($g$) is from right to left. In difference simulation cases, $g$ is adjusted, so that $\beta m g \hat{z}$ varies from 0 to 2. In the ramp surface, from left to right, the component of gravitational acceleration is $\hat{z}g/\hat{L}$. There is no long-range force among the particles. The particle collisions are elastic, calculated by solving Newton's equations.

The left/right borders (AA′ and BB′) and the upper/lower borders (AB and A′B′) are all open, using periodic boundary condition. Initially, the particles are randomly placed in the simulation box, with the probability at height $z$ proportional to the Boltzmann factor, $e^{-\beta m g z}$. The initial particle direction is random. The initial particle speed is randomly assigned, following the 2D Maxwell-Boltzmann distribution $p(v) = mv/(k_\mathrm{B}T_\mathrm{n}) \cdot e^{-mv^2/(2k_\mathrm{B}T_\mathrm{n})}$, where $T_\mathrm{n} = T - E_0/(Nk_\mathrm{B})$, and $E_0$ is the total potential energy of all the particles. The adjustment of $T_\mathrm{n}$ ensures that the expected value of the system energy is the same in all the simulation cases. If the total initial x-component or y-component of momentum of all the particles is larger than 0.1% of $p_0 \triangleq \sqrt{2mk_\mathrm{B}T/\pi}$, or if the total particle energy ($U$) is different from $Nk_\mathrm{B}T$ by more than 0.2%, the randomly generated configuration would be rejected. Thus, the initial particle flow rate is nearly zero, and all the simulation cases have similar $U$.

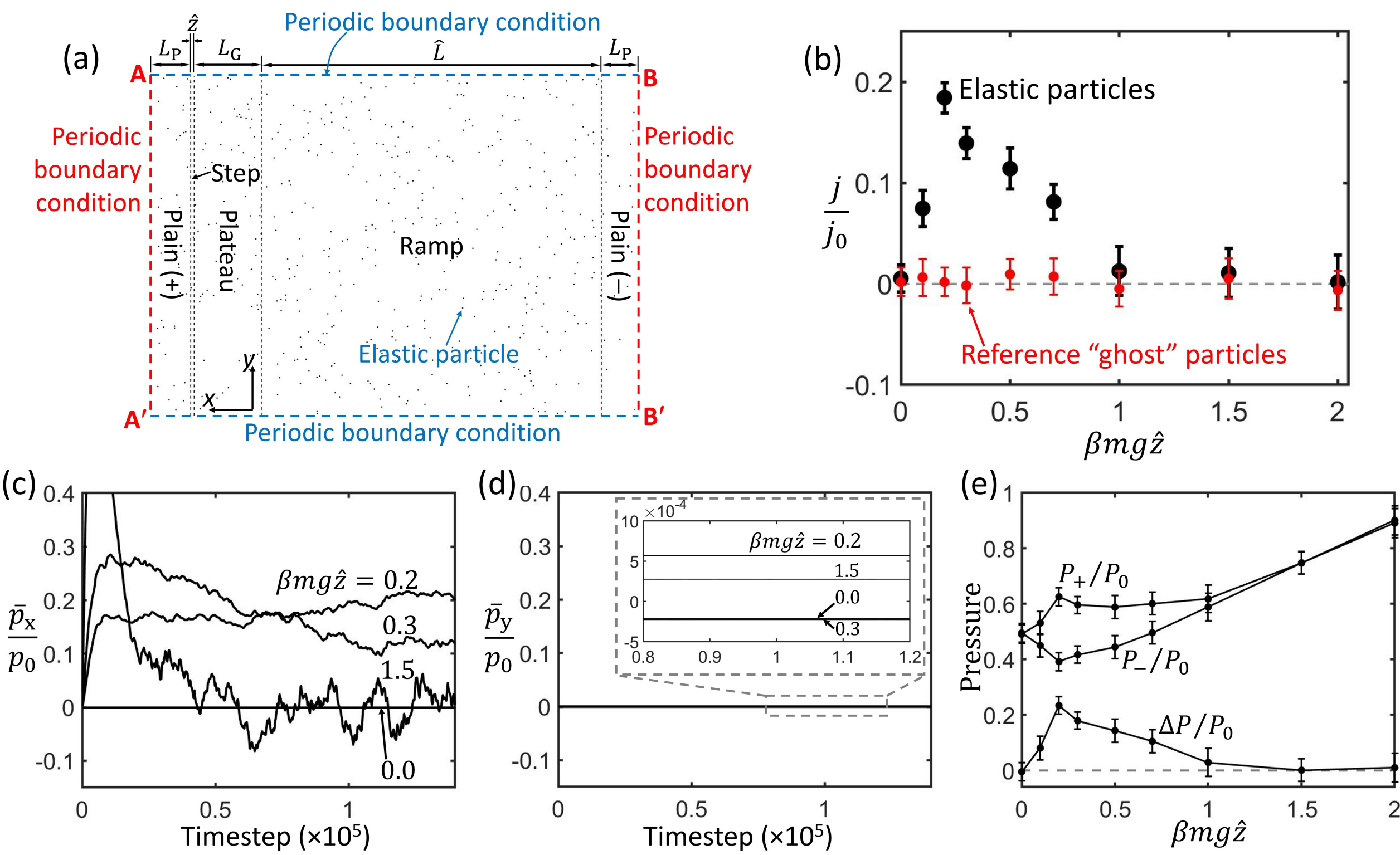

**Fig. 4 (a)** The Monte Carlo simulation. **(b)** The calculated steady-state particle flow rate ($j$) as a function of the normalized energy barrier ($\beta mg\hat{z}$). The red data points show the reference numerical experiment on "ghost" particles, with the particle-particle collision being turned off. **(c)** Typical time profiles of the average x-component of particle momentum ($\bar{p}_\mathrm{x}$) and **(d)** the average y-component of particle momentum ($\bar{p}_\mathrm{y}$); the inset shows a closeup view. **(e)** The inner pressure at the left/right periodic boundary (AA′ and BB′). The normalization constants are defined as $j_0 = \bar{\rho} \cdot \bar{v}_\mathrm{x}/2$, $p_0 = \sqrt{2mk_\mathrm{B}T/\pi}$, and $P_0 = Nk_\mathrm{B}T/A_0$.

Each time a particle crosses the left/right periodic boundary (AA′ and BB′), the time, the speed, and the direction are recorded. The average particle flow rate ($j$) is calculated as $(n_+ - n_-)/(w_0 \Delta t)$ for every $\Delta t = 5000$ timesteps (Figure 4b), where $n_+$ and $n_-$ are the numbers of the crossing events from plain "+" to "–" and from plain "–" to "+", respectively. The total system energy ($E_{\text{tot}}$) is monitored. If the accumulated error in $E_{\text{tot}}$ exceeds 0.1%, the simulation case would be abandoned. Reference tests are performed on "ghost" particles, with the particle collision being turned off; all the other settings remain unchanged. It can be seen that $j \approx 0$ for all the reference cases, suggesting that particle collision is the critical factor.

The average particle momentum is defined as $\bar{p}_{\text{x}} = \frac{1}{N}\sum m v_{\text{x}}$ and $\bar{p}_{\text{y}} = \frac{1}{N}\sum m v_{\text{y}}$, where $\Sigma$ indicates summation for all the particles, and $v_{\text{x}}$ and $v_{\text{y}}$ are the x-component and the y-component of particle velocity, respectively. The time-average $\bar{p}_{\text{x}}$ and $\bar{p}_{\text{y}}$ are computed for every 200 timesteps, as shown in Figure 4(c,d).

The inner pressure is calculated as $P_\pm = \frac{1}{w_0 \Delta t}\sum_\pm m v_{\text{x}}$, where $\Sigma_+$ and $\Sigma_-$ indicate summation in every 5000 timesteps ($\Delta t$) for the particles crossing the left/right periodic boundary (AA′ and BB′) from plain "+" to "–" and from plain "–" to "+", respectively. The overall inner pressure is defined as $\Delta P = P_+ - P_-$ (Figure 4e).

In Section A6 in the Appendix, we perform a number of numerical experiments on different boundary and initial conditions. With the low-height vertical step, similar steady-state particle flows are observed in all the simulation cases.

### 3.3 Monte Carlo simulation: production of useful work

For $\beta m g \hat{z} = 0.2$, after $j$ reaches the steady state, a paddle blade is placed at the middle of plain "+" (Figure 5a). It is modeled as a rigid specular line normal to the x axis, with the length being $w_0$ and the mass ($m_{\text{p}}$) being $200m$, $500m$, or $1000m$. The paddle blade can freely move along the x axis, but does not move along the y axis or rotate.

Figure 5(b,c) shows that the paddle blade is driven by the particle flux. When its displacement ($x_{\text{p}}$) exceeds $0.5L_{\text{p}}$, it crosses the left/right periodic boundary (AA′ and BB′) from plain "+" to "–". Figure 5(d) shows the energy evolution: $U$ is the total kinetic energy and potential energy of all the particles, and the kinetic energy of the paddle blade is $K_{\text{p}} \triangleq m_{\text{p}} v_{\text{p}}^2/2$, where $v_{\text{p}}$ is its speed. The increase in $K_{\text{p}}$ matches the reduction in $U$. The overall energy, $E_{\text{tot}} = U + K_{\text{p}}$,

remains constant. In Section A7 in the Appendix, we show that the movement of the paddle blade is not sensitive to the boundary and initial condition.

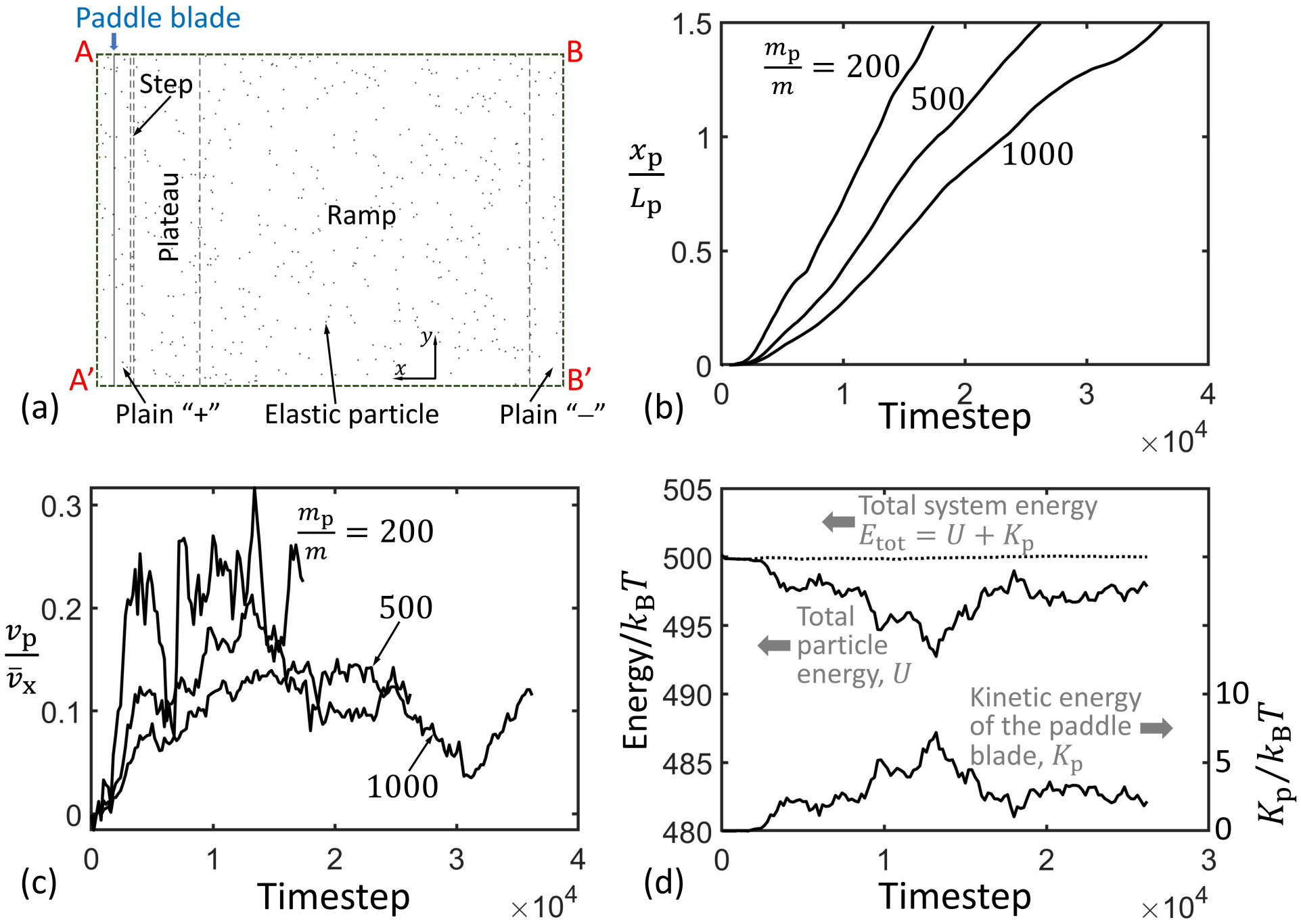

**Fig. 5** **(a)** A paddle blade is driven by the particle flow, converting thermal energy to mechanical energy. **(b)** Typical time profiles of the displacement ($x_p$) and **(c)** the velocity ($v_p$) of the paddle blade. **(d)** Typical energy evolution ($m_p/m = 500$).

Section A8 in the Appendix demonstrates that, from a single thermal reservoir (the heat exchanger), thermal energy can be converted to useful kinetic energy ($K_p$) in a cycle by absorbing heat without any other effect, contradicting the heat-engine statement of the second law of thermodynamics. The energy conversion mechanism is fundamentally different from the Carnot engine. It does not require any temperature difference or fluctuation, and its efficiency is not directly dependent on temperature.

## 4. Discussion

### 4.1 Entropy decrease in the isolated system

Figure 4(b) qualitatively agrees with Figure 3(c). The difference between them should be attributed to the large ramp area, the heterogeneous particle velocity distribution, etc. In both figures, when $\beta mg\hat{z}$ is 0.1~1, a significant particle flux is observed. This is compatible with Figure 4(c-e). Along the y axis, the particle motion is unordered and there is no external force, so that $\bar{p}_{\mathrm{y}}$ remains unchanged (near zero). With the particle flux along +x, the steady-state $\bar{p}_{\mathrm{x}}$ is nontrivial. The change in $\bar{p}_{\mathrm{x}}$ comes from the unbalanced reaction forces on the step and the ramp. Due to the biased particle movement, $P_+ > P_-$ and there is a large $\Delta P$. It serves as the driving force of the paddle blade in Figure 5, converting thermal energy to mechanical energy ($K_{\mathrm{p}}$).

The step-ramp system does not exchange energy or mass with the environment. Its initial state is near equilibrium. A number of observations support that the simulation result reflects the steady state. Firstly, if the system would eventually approach thermodynamic equilibrium, it should not deviate from the equilibrium initial state in the first place. Secondly, as the initial condition is randomized, all the same-setting computational cases demonstrate similar steady-state $j$. If the nonequilibrium state were transient, there is no reason for the simulations to go through the same path. In Figure 11(e) in the Appendix, the system is given various initial x-direction flow rates, with everything else being unchanged; at steady state, $\bar{p}_{\mathrm{x}}$ always converges to the same level. Thirdly, the longest simulation that we have run reached about $4\times10^5$ timesteps (more than 2 times longer than the curves in Figure 4). There was no sign of deviation from the nonequilibrium final state.

The setup in Figure 3(a) can be arbitrarily large and may be analyzed as an ideal gas. If the step and the ramp are much smaller than the plain and the plateau, $N \approx N_{\mathrm{P}} + N_{\mathrm{G}}$, so that $N_{\mathrm{P}} = N/(\hat{\rho}\tilde{A}+1)$ and $N_{\mathrm{G}} = N\hat{\rho}\tilde{A}/(\hat{\rho}\tilde{A}+1)$. When the system is initially at thermodynamic equilibrium, the drift velocity ($v_{\mathrm{w}}$) is zero and the plateau-to-plain particle number density ($\hat{\rho} \triangleq \rho_{\mathrm{G}}/\rho_{\mathrm{P}}$) is the Boltzmann factor ($\delta_{\mathrm{n}}$), and entropy ($S$) reaches the maximum possible value ($S_{\mathrm{eq}}$). Based on the entropy equation of ideal gas [3],

$$S_{\mathrm{eq}} \approx S_{\mathrm{P}} + S_{\mathrm{G}} = N_{\mathrm{P}}k_{\mathrm{B}}\left(\ln\frac{A_{\mathrm{P}}}{N_{\mathrm{P}}} + \sigma_0\right) + N_{\mathrm{G}}k_{\mathrm{B}}\left(\ln\frac{A_{\mathrm{G}}}{N_{\mathrm{G}}} + \sigma_0\right)$$

$$= \frac{Nk_{\mathrm{B}}}{\delta_{\mathrm{n}}\tilde{A}+1}\ln\frac{\delta_{\mathrm{n}}A_{\mathrm{G}}+A_{\mathrm{P}}}{N} + \frac{N\delta_{\mathrm{n}}\tilde{A}k_{\mathrm{B}}}{\delta_{\mathrm{n}}\tilde{A}+1}\ln\frac{\delta_{\mathrm{n}}A_{\mathrm{G}}+A_{\mathrm{P}}}{N\delta_{\mathrm{n}}} + Nk_{\mathrm{B}}\sigma_0 \tag{4}$$

where $S_{\mathrm{P}}$ is the entropy of the plain, $S_{\mathrm{G}}$ is the entropy of the plateau, $\delta_{\mathrm{n}} = e^{-mg\hat{z}/(k_{\mathrm{B}}T_{\mathrm{n}})}$, and $\sigma_0 = \ln(2\pi emk_{\mathrm{B}}T_{\mathrm{n}})$. It can be verified that, with $N_{\mathrm{P}} = N - N_{\mathrm{G}}$ and $k_{\mathrm{B}}T_{\mathrm{n}} = (U - N_{\mathrm{G}}mg\hat{z})/N$, when $\hat{\rho} = \delta_{\mathrm{n}}$, $\frac{\partial S_{\mathrm{eq}}}{\partial N_{\mathrm{G}}} = -k_{\mathrm{B}}\left(\ln\delta_{\mathrm{n}} + \frac{mg\hat{z}}{k_{\mathrm{B}}T_{\mathrm{n}}}\right) = 0$, so that $S_{\mathrm{eq}}$ is maximized. At the nonequilibrium steady state, $\hat{\rho} \approx \bar{\delta}$ and $v_{\mathrm{w}} > 0$, and $S$ converges to the nonequilibrium value ($S_{\mathrm{ne}}$) [15,16], which may be estimated similarly to Equation (4):

$$S_{\mathrm{ne}} \approx S_{\mathrm{P}} + S_{\mathrm{G}} \approx N_{\mathrm{P}} k_{\mathrm{B}} \left( \ln \frac{A_{\mathrm{P}}}{N_{\mathrm{P}}} + \sigma \right) + N_{\mathrm{G}} k_{\mathrm{B}} \left( \ln \frac{A_{\mathrm{G}}}{N_{\mathrm{G}}} + \sigma \right)$$

$$\approx \frac{N k_{\mathrm{B}}}{\bar{\delta}\tilde{A}+1} \ln \frac{\bar{\delta} A_{\mathrm{G}} + A_{\mathrm{P}}}{N} + \frac{N \bar{\delta} \tilde{A} k_{\mathrm{B}}}{\bar{\delta}\tilde{A}+1} \ln \frac{\bar{\delta} A_{\mathrm{G}} + A_{\mathrm{P}}}{N \bar{\delta}} + N k_{\mathrm{B}} \sigma \tag{5}$$

where $\sigma = \ln\left(2\pi e m k_{\mathrm{B}} \hat{T}\right)$ and $\hat{T} = T_{\mathrm{n}} - m v_{\mathrm{w}}^2/(2k_{\mathrm{B}})$.

Generally, $S_{\mathrm{ne}} < S_{\mathrm{eq}}$. As the ordered particle flow is formed, $S$ is reduced from $S_{\mathrm{eq}}$ to $S_{\mathrm{ne}}$; i.e., $S \to S_{\mathrm{Q}}$. The entropy decrease is $\Delta S = S_{\mathrm{eq}} - S_{\mathrm{ne}}$. For example, by using the parameters in Section 3.2, for $\beta m g \hat{z} = 0.2$, Equations (4) and (5) suggest that $S_{\mathrm{eq}} = 6.14 \times 10^3$ and $S_{\mathrm{ne}} = 6.09 \times 10^3$; the corresponding $T_{\mathrm{n}} \Delta S$ is about 10% of $N k_{\mathrm{B}} T$. The entropy decrease can be attributed to the difference of $\bar{\delta}$ and $\hat{T}$ from $\delta_0$ and $T_{\mathrm{n}}$. The former ($\bar{\delta}$) represents the effects of $\hat{\rho}$; the latter ($\hat{T}$) represents the degree of randomness of $v$. Equation (5) assumes that the distribution of particle velocity is homogeneous. If we take into account the heterogeneity of $v$ (see Section 4.3), the calculated $S_{\mathrm{ne}}$ would be even smaller, as the system is more nonuniform.

In Section 5.2 below, the reduction in entropy is interpreted by using the method of Lagrange multipliers:

$$\frac{\partial \mathcal{L}}{\partial n_i} = 0 \tag{6}$$

where $\mathcal{L}$ is the Lagrangian, and $n_i$ is the particle number at the $i$-th energy level ($\epsilon_i$). The nonchaotic particle movement in the narrow energy barrier imposes additional constraints on $n_i$, so that entropy is maximized to the more constrained nonequilibrium maximum ($S_{\mathrm{ne}}$), lower than the less constrained equilibrium maximum ($S_{\mathrm{eq}}$). In this framework, $\Delta S$ is consistent with the basic principle of maximum entropy.

As the steady state is nonequilibrium, there are a variety of possible initial states of higher entropy than $S_{\mathrm{Q}}$. In addition to the near-equilibrium initial state in Section 3.2, another example is the state that has similar particle number density and particle speed distributions to the steady state, but no particle flux (i.e., initially $v_{\mathrm{w}} = 0$). It may be realized by blocking the step and the ramp, so that the plain and the plateau are thermalized separately. After the step and the ramp are opened, as $v_{\mathrm{w}}$ is built up, $S$ decreases and converges to $S_{\mathrm{ne}}$.

Since the steady-state particle flow is continuous, it does not consume energy from the gravitational field. On average, corresponding to every ascending particle in the ramp, there is a descending particle in the step; vice versa. The produced mechanical energy ($K_{\mathrm{p}}$) in Section 3.3 is from thermal energy (see Figure 5d).

4.2 Isolation and the intrinsically nonequilibrium steady state

It has long been known that certain "peculiar" systems cannot reach thermodynamic equilibrium and should not be analyzed by thermodynamics, such as some nonergodic or nonchaotic particle movements [23-25]. Traditionally, people do not consider them as a violation to thermodynamics, because the system size is small, the energy properties are "trivial", and/or thermodynamic equilibrium is not accessible.

For a large-sized "regular" system (e.g., an ideal gas), the second law of thermodynamics dictates that, without an energetic penalty, the steady state cannot be nonequilibrium. For instance, in an isolated setup, the steady-state gas pressure across a porous membrane must be uniform, regardless of the pore size and pore shape [26]. Otherwise, it would cause a "Maxwell's demon type" controversy [16].

Yet, Section 3 demonstrates a counterexample. The entropy decrease discussed in Section 4.1 and the thermal-to-mechanical energy conversion shown in Figure 5 are the result of the intrinsically nonequilibrium steady state, caused by the lack of particle-particle collision in the SND. Local nonchaoticity renders the H-theorem inapplicable, so that there is no mechanism to drive entropy to increase to $S_{\mathrm{eq}}$.

As long as the steady state is significantly different from thermodynamic equilibrium (e.g., $\hat{\rho} \neq \delta_{\mathrm{n}}$), the step and the ramp may be arranged in a variety of ways to harvest thermal energy from a single heat reservoir, not necessarily through a particle flow. Section A9 in the Appendix demonstrates such an example, in which the step and the ramp are alternately opened and closed, and a frictionless piston cyclically expands and compresses the plateau.

4.3 Nonuniform distribution of particle velocity

In a gravitational field, temperature of an isolated chaotic gas is uniform along height [27,28]. Maxwell stated [29]: "…If the temperature of any substance, when in thermal equilibrium, is a function of height, that of any other substance must be the same function of the height. For if not, let equal columns of the two substances be enclosed in cylinders impermeable to heat, and put in thermal communication at the bottom. If, when in thermal equilibrium, the tops of the two columns are at different temperatures, an engine might be worked by taking heat from the hotter and giving it up to the cooler, and the refuse heat would circulate round the system till it was all converted into mechanical energy, which is contradiction to the second law of thermodynamics. The result as now given is, that temperature in gases, when in thermal equilibrium, is independent of height, and it follows from what has been said that temperature is independent of height in all other substances." Hence, in the wide ramp in Figure 3(a), at steady state, the particle velocity distribution tends to be homogenous.

Across the vertical step, as particle-particle collision is lacking, there is no mechanism to ensure thermal equilibrium. In the numerical simulation in [15], we observed that when $\beta m g \hat{z} = 0.5$, at the steady state, the average particle speed on the plateau was greater than on the plain by ~10%. It is nontrivial, but relatively less important compared to the strong nonchaoticity effect of the particle flux ratio. For example, under the same condition of $\beta m g \hat{z} = 0.5$, $\delta_0$ ($\approx 0.607$) is larger than $\delta_1$ ($\approx 0.317$) by nearly two times. The heterogeneous particle velocity distribution could result in unusual phenomena of heat transfer [30] and might affect the particle flow. These will be important topics for future study.

In Equation (5), if the difference in the average particle kinetic energy ($\overline{K}$) between the plateau and the plain is taken into consideration, the expression of $\sigma$ needs to be modified. The calculated $S_{\mathrm{ne}}$ would be even lower, as the degree of nonuniformity increases.

### 4.4 Considerations on experimental realization

If the particles are air molecules, at room temperature, $\bar{v}_{\mathrm{x}}$ is ~270 m/s. According to Equation (3), the maximum drift velocity is 50~60 m/s, comparable with the wind speed of a Category 5 hurricane. Yet, as mentioned in Section 2.1, to achieve a nontrivial particle flow, $g$ must be greater than $10^{11}$ m/s$^2$, over 10 orders of magnitude stronger than the gravitational acceleration on Earth, ~4 orders of magnitude beyond the capacity of centrifuges. Using heavy gas molecules with a powerful centrifuge at a low temperature might help to overcome this hurdle. A more straightforward approach is probably to use a stronger thermodynamic force, such as the Coulomb force. To work with a Coulomb force, the particles should be charged, e.g., the dissolved ions in an electrolyte solution, the charge carriers in a Fermi gas or a plasma, etc. Section A10 in the Appendix presents some considerations on the charge carriers, suggesting that the upper limit of the power density could be more than 10 $kW/cm^3$.

In addition to the gravitational/inertia force and the Coulomb force, there are other relevant thermodynamic forces, such as magnetic force, gas/plasma pressure, degeneracy pressure, and chemical potential. An important future research topic is to explore whether SND-like phenomena may exist in nature, e.g., on atomic and molecular scales, on subatomic scales, in high-$g$ environments, or for weakly/sparsely interacting particles. The possible nonequilibrium mechanisms and the associated state evolution in phase space should be investigated in the context of classical mechanics, relativistic mechanics, and quantum mechanics.

Notice that SND does not have to be an energy barrier. It could also be a locally nonchaotic entropy barrier [16]. The concept of entropy-barrier SND has been experimentally realized by using nanoporous membranes one-sidedly surface-treated with bendable organic chains [16].

## 5. Extended Discussion: A Variant System with a Stepped Plain-Plateau Border

This section presents a reanalysis of the numerical findings reported in [15], which might be helpful for further understanding the step-ramp model in Sections 3-4. It offers another perspective to view the global consequence of local nonchaoticity, without the need to deal with the particle flow. Sections 5.2 and 5.4 are new.

### 5.1 System design and operation: production of useful work in an isothermal cycle

Figure 6(a) depicts the ideal-gas model analyzed in [15]. It consists of a plateau and a plain. Different from Figure 3(a), the entire plateau-plain boundary is a vertical step. In [15], it has been demonstrated that when the step height ($\hat{z}$) is much less than the nominal particle mean free path ($\lambda_F$), because of the intrinsically nonequilibrium particle flux ratio ($\delta_*$), the steady-state plateau-to-plain particle number density ratio ($\hat{\rho} \triangleq \rho_G/\rho_P$) is considerably less than $\delta_0$ (Figure 6b).

The system is closed. The outer plain boundary is in contact with a thermal bath. The plateau height ($\hat{z}$) and the plain area ($A_P$) are adjustable, and can be operated in the four-step isothermal cycle in Figure 6(c,d). The plateau is first raised by the support force ($F_G$) from $\hat{z}_L$ to $\hat{z}_u$ (State I to II); then, the plain area is expanded by the in-plane pressure ($P$) from $A_{Pl}$ to $A_{Pu}$ (State II to III), followed by the decrease of $\hat{z}$ back to $\hat{z}_L$ (State III to IV); finally, the plain is compressed by $P$ to $A_{Pl}$ (State IV to I).

In general, in an equilibrium system with two thermally correlated thermodynamic forces ($F_a$ and $F_b$), because $F_a = \frac{\partial \mathcal{A}}{\partial x_a}$ and $F_b = \frac{\partial \mathcal{A}}{\partial x_b}$,

$$\frac{\partial F_a}{\partial x_b} = \frac{\partial F_b}{\partial x_a} \tag{7}$$

where $\mathcal{A} = U - TS$ is the Helmholtz free energy, $U$ is the internal energy, and $x_a$ and $x_b$ are the conjugate variables of $F_a$ and $F_b$, respectively. Examples of Equation (7) include the Maxwell relations [31], the Nernst equation [32], the Lippman equation [33], the relationship between the surface tension and the electrolyte concentration [33], etc. It reflects the heat-engine statement of the second law of thermodynamics [15] (see Section A11 in the Appendix). In Figure 6(a), we assume that the step area is much smaller than the plain area and the plateau area. The kinetic analysis indicates that $F_G = mgN_G$ and $P = N_P\overline{K}/A_P$ [15]. The conjugate variables of $F_G$ and $P$ are $\hat{z}$ and $-A_P$, respectively. For $F_G$ and $P$, Equation (7) becomes $-\frac{\partial F_G}{\partial A_P} = \frac{\partial P}{\partial \hat{z}}$, which can be rewritten as $\frac{\partial \hat{\rho}}{\partial \hat{z}} = -\beta mg\hat{\rho}$. Its solution, $\hat{\rho} = e^{-\beta mg\hat{z}}$, is the Maxwell-Boltzmann distribution. Hence, at equilibrium ($\hat{\rho} = \delta_0$), in the 4-step isothermal cycle, the total work generated by the in-plane pressure $P$ ($W_P$) equals to the total work consumed by the support force $F_G$ ($W_G$).

Yet, at the nonequilibrium steady state, since $\hat{\rho} \neq e^{-\beta mg\hat{z}}$ (Figure 6b), $\frac{\partial \hat{\rho}}{\partial \hat{z}} \neq -\beta mg\hat{\rho}$, so that Equation (7) cannot be balanced. As a result, in the isothermal cycle, the produced work ($W_P$) is greater than the consumed work ($W_G$) [15], conflicting with the second law of thermodynamics. When $\hat{\rho} \approx \delta_1$, $W_P > W_G$ can be observed by calculating $W_G = mgN \cdot \int_{\hat{z}_L}^{\hat{z}_u} \left\{ \left[1 + 1/\left(\tilde{A}_I\hat{\rho}\right)\right]^{-1} - \left[1 + 1/\left(\tilde{A}_{II}\hat{\rho}\right)\right]^{-1} \right\} \mathrm{d}\hat{z}$ and $W_P = N\overline{K} \cdot \ln(A_U/A_L)$, where $\tilde{A}_I = A_G/A_{Pl}$, $\tilde{A}_{II} = A_G/A_{Pu}$, $A_U =$

$(A_{\mathrm{Pu}} + A_{\mathrm{G}}\delta_{\mathrm{u}})(A_{\mathrm{Pl}} + A_{\mathrm{G}}\delta_{\mathrm{L}})$, $A_{\mathrm{L}} = (A_{\mathrm{Pl}} + A_{\mathrm{G}}\delta_{\mathrm{u}})(A_{\mathrm{Pu}} + A_{\mathrm{G}}\delta_{\mathrm{L}})$, $\delta_{\mathrm{u}} = 1 - \mathrm{erf}\left(\sqrt{\beta m g \hat{z}_{\mathrm{u}}}\right)$, and $\delta_{\mathrm{L}} = 1 - \mathrm{erf}\left(\sqrt{\beta m g \hat{z}_{\mathrm{L}}}\right)$.

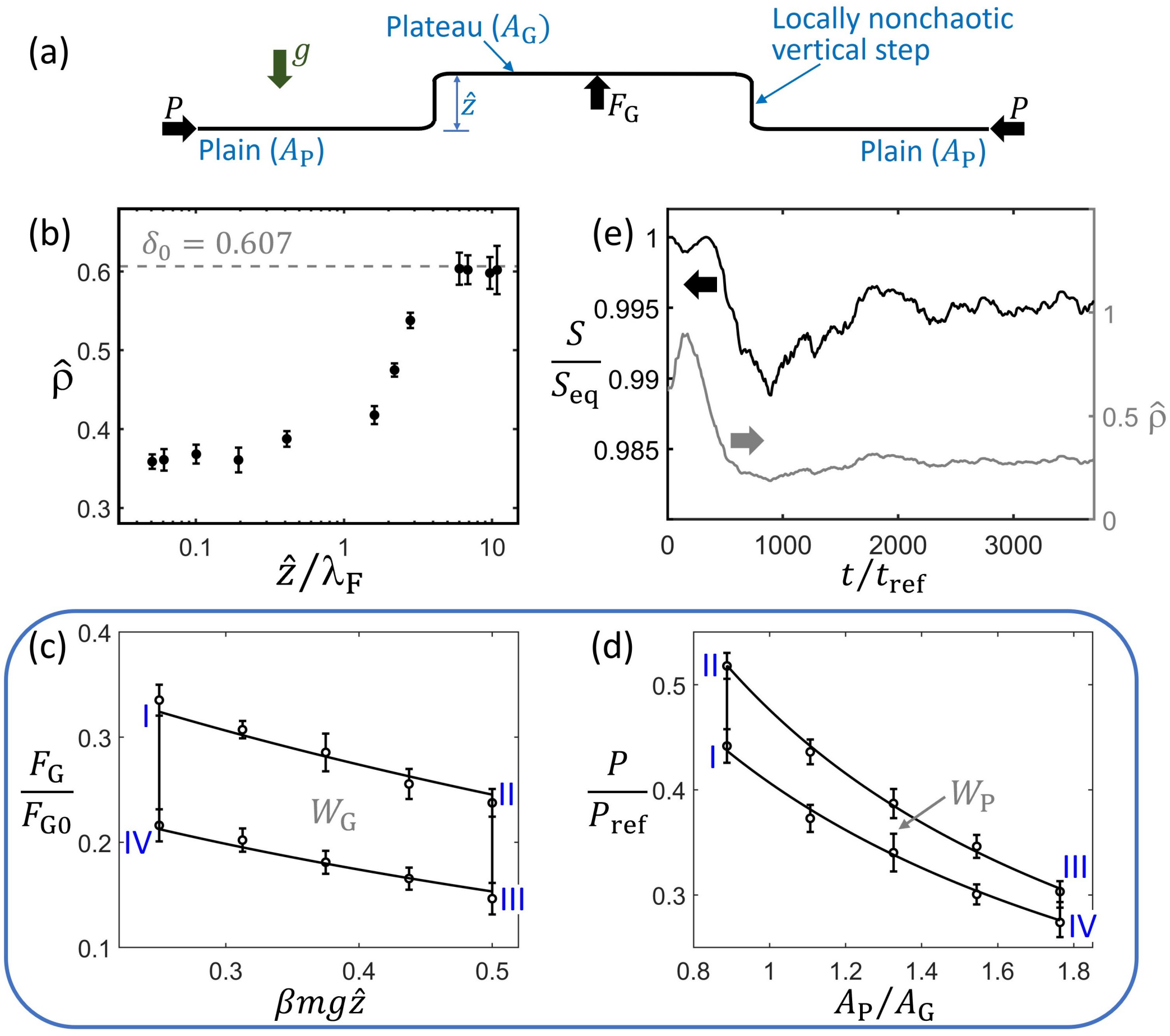

**Fig. 6 (a)** Two large horizontal areas at different heights (the upper plateau and the lower plain) are connected by a vertical step [15], in which 2D elastic particles randomly move. A uniform gravitational field ($g$) is along the vertical direction. The MC simulation results: **(b)** The steady-state plateau-to-plain particle number density ratio ($\hat{\rho}$) as a function of $\hat{z}/\lambda_{\mathrm{F}}$. When $\hat{z}/\lambda_{\mathrm{F}} \ll 1$, the step is locally nonchaotic and $\hat{\rho}$ is much smaller than the Boltzmann factor ($\delta_0$). **(c)** In a 4-step isothermal cycle ($\hat{z}/\lambda_{\mathrm{F}} \approx 0.1$), the system is changed from State I to II, III, IV, and back to I. The operation of $F_{\mathrm{G}}$ consumes work ($W_{\mathrm{G}}$), and **(d)** the operation of $P$ produces work ($W_{\mathrm{P}}$). The normalization constants are $F_{\mathrm{G0}} = mgN$ and $P_{\mathrm{ref}} = Nk_{\mathrm{B}}T/A_{\mathrm{G}}$. As a result of the nonequilibrium steady state ($\hat{\rho} < \delta_0$), $W_{\mathrm{P}}$ is significantly larger than $W_{\mathrm{G}}$ ($W_{\mathrm{P}}/W_{\mathrm{G}} = 1.704$). **(e)** Typical time profile of the entropy decrease process in an isolated setup. The normalization constant of time ($t$) is $t_{\mathrm{ref}} = \hat{z}\sqrt{\beta m/2} = 0.245$; $\delta_0 = 0.607$; $S_{\mathrm{eq}} = 1.038 \times 10^4$; $\hat{z}/\lambda_{\mathrm{F}} \approx 0.0225$.

The difference between $W_{\mathrm{P}}$ and $W_{\mathrm{G}}$ may be viewed through the overall system governing equations, $PA_0 = \varepsilon_{\mathrm{P}} N k_{\mathrm{B}} T$ and $F_{\mathrm{G}} = \varepsilon_{\mathrm{G}} mgN$, where $A_0 = A_{\mathrm{G}} + A_{\mathrm{P}}$, $\varepsilon_{\mathrm{P}} = A_0/(A_{\mathrm{P}} + \hat{\rho}A_{\mathrm{G}})$, and $\varepsilon_{\mathrm{G}} = 1/[1 + A_{\mathrm{P}}/(\hat{\rho}A_{\mathrm{G}})]$. If the plateau height is zero, $\hat{\rho} = 1$, so that $\varepsilon_{\mathrm{P}} = 1$ and $\varepsilon_{\mathrm{G}} = A_{\mathrm{G}}/A_0$;

thus, the equations are reduced to $PA_0 = Nk_\mathrm{B}T$ and $F_\mathrm{G} = mgNA_\mathrm{G}/A_0$. With a nonzero $\hat{z}$, at thermodynamic equilibrium, $\hat{\rho} = \delta_0$ and $W_\mathrm{P} = W_\mathrm{G}$. At the nonequilibrium steady state, because $\hat{\rho} < \delta_0$, compared to the equilibrium state, there are less particles on the plateau and more particles in the plain. As $P$ is larger than the equilibrium pressure while $F_\mathrm{G}$ is smaller than the equilibrium support force, $W_\mathrm{P}$ tends to be greater than $W_\mathrm{G}$.

The MC simulation in Figure 6(c,d) confirms the above analysis ($\hat{z}/\lambda_\mathrm{F} \approx 0.1$). The numerical procedure is detailed in [15]. In Figure 6(c), the solid regression curves are $F_\mathrm{G} \approx mgN_\mathrm{G} = mgN\rho_\mathrm{z}A_\mathrm{G}/(\rho_\mathrm{z}A_\mathrm{G} + A_\mathrm{P})$, where $\rho_\mathrm{z} = \alpha_\mathrm{n}\left[1 - \mathrm{erf}\left(\sqrt{\beta mg\hat{z}}\right)\right]$ and $\alpha_\mathrm{n}$ is an adjustable parameter. For the upper curve, $\alpha_\mathrm{n}$ is set to 1.095; for the lower curve, $\alpha_\mathrm{n}$ is set to 1.185. In Figure 6(d), the solid regression curves are formulated as $P \approx N_\mathrm{P}\bar{K}/A_\mathrm{P} = N\bar{K}/(A_\mathrm{P} + A_\mathrm{G}\hat{\rho})$, with the average $\hat{\rho}$ and the effective $\bar{K}$ being computed from the simulation data [15]. The consumed and produced works ($W_\mathrm{G}$ and $W_\mathrm{P}$) are assessed as the areas enclosed in between the upper and lower curves in Figure 6(c) and Figure 6(d), respectively: $W_\mathrm{G} = 22.81k_\mathrm{B}T$ and $W_\mathrm{P} = 34.61k_\mathrm{B}T$; $W_\mathrm{P}/W_\mathrm{G} = 1.704$ is significantly larger than unity.

### 5.2 Entropy decrease in an isolated setup

Figure 6(e) shows the time profile of an entropy decrease process, when the system is isolated. Initially, the particles are randomly placed on the plain and the plateau, with the probability on the plateau less than on the plain by a factor of $\delta_0$, i.e., the initial $\hat{\rho} = \delta_0$. The initial particle direction is random; the initial particle speed randomly follows $p(v)$ ($T = 1000$). The plain is circular, with a radius of 400. The outer plain border is a rigid diffusive wall; the reflected direction is random; the reflected speed is equal to the incident speed. The plateau is a circular area at the center, with a radius of 200. The normalized energy barrier is $\beta mg\hat{z} = 0.5$ ($\delta_0 = 0.607$; $\delta_1 = 0.317$); $N = 1000$; $\hat{z} = 1$; $d = 4$; $m = 1$; $\Delta t_0 = 0.001$; $\hat{z}/\lambda_\mathrm{F} \approx 0.0225$. The simulation is scalable; an example of the unit system can be based on Å, ps, g/mol, and K. The numerical procedure is similar to that of Figure 6(b) [15]. The computer program is available at [17].

Entropy is calculated similarly to Equation (4) [3]: $S = S_\mathrm{P} + S_\mathrm{G} = k_\mathrm{B}N_\mathrm{P}[\ln(A_\mathrm{P}/N_\mathrm{P}) + \sigma_0] + k_\mathrm{B}N_\mathrm{G}[\ln(A_\mathrm{G}/N_\mathrm{G}) + \sigma_0]$. The system spontaneously deviates from the near-equilibrium initial state ($\hat{\rho} \approx \delta_0$) and converges to the nonequilibrium steady state ($\hat{\rho} < \delta_0$). Accordingly, $S$ decreases from $S_{\mathrm{eq}}$ to $S_{\mathrm{ne}}$ ($S_{\mathrm{ne}}/S_{\mathrm{eq}} \approx 0.995$).

The reduction in entropy may be understood by using the method of Lagrange multipliers [34]. For an isolated chaotic system ($\hat{z} \gg \lambda_{\mathrm{F}}$), to maximize the total number of microstates ($\Omega$), under the constraints of $\sum_i n_i = N$ and $\sum_i n_i \epsilon_i = U$, the Lagrangian can be written as $\mathcal{L}_0 = \ln \Omega + \lambda_1 (N - \sum_i n_i) + \bar{\beta}_1 (U - \sum_i n_i \epsilon_i)$, where $\lambda_1$ and $\bar{\beta}_1$ are the Lagrange multipliers. When the system is dilute and the density of states $g_i \gg n_i$, $\ln \Omega \approx \sum_i (n_i \ln g_i - n_i \ln n_i + n_i)$. In accordance with $\frac{\partial \mathcal{L}_0}{\partial n_i} = 0$ (Equation 6), $n_i \propto g_i e^{-\beta \epsilon_i}$. Since $g_i$ is proportional to the area of particle movement, $\hat{\rho} \triangleq \rho_{\mathrm{G}}/\rho_{\mathrm{P}} = \int_{mg\hat{z}}^{\infty} (n_i/g_i) \mathrm{d}\epsilon_i / \int_0^{\infty} (n_i/g_i) \mathrm{d}\epsilon_i = e^{-\beta m g \hat{z}}$. It leads to the maximum possible entropy at thermodynamic equilibrium, $S_{\mathrm{eq}}$ (Equation 4).

When $\hat{z} \ll \lambda_{\mathrm{F}}$, compared to the fully chaotic setup, the microstates are less random. The "deterministic" particle transmission in the vertical step imposes an additional constraint: $\hat{\rho} = \delta_*$. If $\delta_* = \delta_0$, the constraint of $\hat{\rho} = \delta_*$ is trivial, because as shown above, it can be derived from the conventional constraints ($\sum_i n_i = N$ and $\sum_i n_i \epsilon_i = U$). When $\delta_* \neq \delta_0$, at thermal equilibrium, $\hat{\rho} = \delta_*$ may be expressed in the strong form: $\frac{(n_i/g_i)_{\mathrm{G}}}{(n_i/g_i)_{\mathrm{P}}} = \xi_1$, where $\xi_1 \approx \delta_1/\delta_0$, and subscripts "G" and "P" indicate the plateau and the plain, respectively. Redefine the Lagrangian as $\mathcal{L}_1 = -\sum_\chi f_\chi \ln f_\chi + \lambda_1 (N - \sum_i n_i) + \bar{\beta}_1 (U - \sum_i n_i \epsilon_i) + \sum_i \{ \tilde{\lambda}_i [ (n_i/g_i)_{\mathrm{G}} - \xi_1 (n_i/g_i)_{\mathrm{P}} ] \}$, with $f_\chi$ being the probability of the $\chi$-th microstate, and $\tilde{\lambda}_i$ the Lagrange multipliers. Following the principle of maximum entropy (Equation 6), we have $\frac{\partial \mathcal{L}_1}{\partial (n_i)_{\mathrm{v}}} = 0$, where subscript "v" represents either "G" or "P". With $-\sum_\chi f_\chi \ln f_\chi = \ln \Omega$, it can be derived that $(n_i)_{\mathrm{G}} = g_i e^{-\bar{\beta} \epsilon_i + \tilde{\lambda}_i / g_i - \lambda_1}$ and $(n_i)_{\mathrm{P}} = g_i e^{-\bar{\beta} \epsilon_i - \tilde{\lambda}_i \xi_1 / g_i - \lambda_1}$, which are clearly non-Boltzmannian. The more constrained maximization using $\mathcal{L}_1$ gives a smaller entropy ($S_{\mathrm{ne}}$) than the less constrained maximization using $\mathcal{L}_0$. It is worth noting that because $\hat{\rho} \neq \delta_0$, Boltzmann's assumption of equal *a priori* equilibrium probabilities [3] may need to be reanalyzed. Moreover, the temperature field is heterogeneous across the vertical step (see Section 4.3). As the actual system tends to be more nonuniform, the calculated $S_{\mathrm{ne}}$ would be further reduced. In any case, $S_{\mathrm{ne}} < S_{\mathrm{eq}}$.

In addition to $\hat{\rho} = \delta_*$, the SND-induced constraint may be expressed in various other forms. Section A12 in the Appendix presents one such example: $f_{n_1}/f_{n_2} = \delta_*^{\Delta n_{12}}$, where $f_{n_1}$ and $f_{n_2}$ are the probabilities of two microstates having the same $\overline{K}$, and $\Delta n_{12}$ is the difference in $N_\mathrm{G}$ between them. Like $\hat{\rho} = \delta_*$, it leads to the more constrained maximation of entropy, rendering $S_\mathrm{ne}$ smaller than $S_\mathrm{eq}$.

5.3 Considerations about local equilibrium, free energy, and the Clausius theorem

Previously in [15], the nonequilibrium $\hat{\rho}$ was formulated as $\delta_0^{\hat{\alpha}}$, with $\hat{\alpha}$ being an adjustable parameter. When $\hat{z}/\lambda_\mathrm{F} \approx 0.1$, the numerical data of $\hat{\rho}$ were in between $\delta_0$ and $\delta_1$, approximately $\delta_0^2$, i.e., $\hat{\alpha} \approx 2$. The exponential form ($\delta_0^{\hat{\alpha}}$) originally came from the hypothesis of local equilibrium [3, 35-37] and also a mistake in the integration of $p_\mathrm{z}$; it did not affect the numerical simulation. In current research, with the narrow energy barrier, $\delta_0^{\hat{\alpha}}$ is no longer used.

The nonequilibrium steady state cannot be directly governed by any thermodynamic free energy, such as the Helmholtz free energy, $\mathcal{A} = U - TS$ [15]. This is the reason why, for the SND-based model system, Equation (7) is irrelevant. In Figure 6(a), only when $\hat{\rho} = \delta_0$, $F_\mathrm{G} \triangleq mgN_\mathrm{G} = \frac{\partial \mathcal{A}}{\partial \hat{z}}$ and $P \triangleq N_\mathrm{P} k_\mathrm{B} T / A_\mathrm{P} = -\frac{\partial \mathcal{A}}{\partial A_\mathrm{P}}$ [15]. If $\hat{\rho} \approx \delta_1$,

$$F_\mathrm{G} = \frac{\partial \mathcal{A}}{\partial \hat{z}} + \sqrt{\frac{\beta m g}{\pi \hat{z}}} \frac{N \tilde{A} \delta_0 \Theta}{(1+\delta_1 \tilde{A})^2} \tag{8}$$

$$P = -\frac{\partial \mathcal{A}}{\partial A_\mathrm{P}} - \frac{N \tilde{A} \delta_1 \Theta}{A_\mathrm{P}(1+\delta_1 \tilde{A})^2}$$

where $\Theta = mg\hat{z} + k_\mathrm{B} T \ln \delta_1$. The second terms at the right-hand side of Equation (8) are caused by the nonequilibrium effect. Nevertheless, local Helmholtz free energy may be separately defined for the plain ($\mathcal{A}_\mathrm{P}$) and the plateau ($\mathcal{A}_\mathrm{G}$), excluding the SND in between them. With a uniform $T$, for the plain, $\mathcal{A}_\mathrm{P} = N_\mathrm{P} k_\mathrm{B} T - T S_\mathrm{P}$; for the plateau, $\mathcal{A}_\mathrm{G} = N_\mathrm{G} k_\mathrm{B} T + N_\mathrm{G} m g \hat{z} - T S_\mathrm{G}$, where $S_\mathrm{P} = N_\mathrm{P} k_\mathrm{B} \cdot [\ln(A_\mathrm{P}/N_\mathrm{P}) + \overline{\sigma}_1]$, $S_\mathrm{G} = N_\mathrm{G} k_\mathrm{B} [\ln(A_\mathrm{G}/N_\mathrm{G}) + \overline{\sigma}_1]$, and $\overline{\sigma}_1 = \ln(2\pi e m k_\mathrm{B} T)$. It can be verified that $-\frac{\partial \mathcal{A}_\mathrm{P}}{\partial A_\mathrm{P}} = \frac{N_\mathrm{P} k_\mathrm{B} T}{A_\mathrm{P}} = P$ and $\frac{\partial \mathcal{A}_\mathrm{G}}{\partial \hat{z}} = N_\mathrm{G} m g = F_\mathrm{G}$. The SND effect could be treated as the boundary condition between the plain and the plateau.

Both entropy and thermal energy are extensive, so that $S = S_\mathrm{P} + S_\mathrm{G}$ and $Q = Q_\mathrm{P} + Q_\mathrm{G}$, where $Q$, $Q_\mathrm{P}$, and $Q_\mathrm{G}$ are the absorbed heat of the system, the plain, and the plateau, respectively.

If temperature ($T$) were uniform in the system, for a reversible process, the change in $S_\mathrm{P}$ is $Q_\mathrm{P}/T$ and the change in $S_\mathrm{G}$ is $Q_\mathrm{G}/T$. Thus, $\mathrm{d}S = Q/T$, where $\mathrm{d}S$ is the change in $S$. Since the local temperature may be different on the plateau and the plain (see Section 4.3), the relationship of $\mathrm{d}S = Q/T$ needs to be revisited.

## 5.4 A variant system with a wide plain-plateau transition zone

Figure 7 depicts a variant system of Figure 6(a), in which the SND elements are not at the boundary but in the interior of the plateau. The transition zone between the plain and the plateau is a wide ramp ($\hat{L} \gg \lambda_\mathrm{F}$). On the plateau, there are a number of vertical-walled stages. Each stage is connected to the plateau through a two-layer gear and a set of racks. Denote the radius ratio of the gear layers by $\zeta_0$ ($\zeta_0 > 1$). It keeps the stage displacement larger than the plateau displacement by a factor of $\zeta_0$. The stage height is $h_\mathrm{t} = \zeta_0 \hat{z}$. The height difference between the stage and the plateau is $z_\mathrm{t} = \zeta_1 \hat{z}$, where $\zeta_1 = \zeta_0 - 1$. Assume that the ramp area and the stage-wall area are much smaller than the plateau area ($A_\mathrm{G}$), the plain area ($A_\mathrm{P}$), and the total stage area ($A_\mathrm{T}$).

Since the particle movement in the ramp is chaotic, $\rho_\mathrm{G} = \rho_\mathrm{P}\delta_0$. To adjust $\hat{z}$ and $h_\mathrm{t}$, the required support force is $F_\mathrm{G} = mgN_\mathrm{G} + F_\mathrm{T} = mg\left(\delta_0 A_\mathrm{G} + \zeta_0 \delta_0 \delta_\mathrm{S} A_\mathrm{T}\right)\rho_\mathrm{P}$, where $F_\mathrm{T} = \zeta_0 mgN_\mathrm{T}$, $\delta_\mathrm{S} = \rho_\mathrm{S}/\rho_\mathrm{G}$, $\rho_\mathrm{S}$ is the particle number density on the stages, and $N_\mathrm{T} = \rho_\mathrm{S} A_\mathrm{T}$. With $\rho_\mathrm{P} = N/(A_\mathrm{P} + \delta_0 A_\mathrm{G} + \delta_0 \delta_\mathrm{S} A_\mathrm{T})$, it can be verified that $\frac{\partial P}{\partial \hat{z}} = \frac{1}{\beta}\frac{\partial \rho_\mathrm{P}}{\partial \hat{z}}$ is equal to $-\frac{\partial F_\mathrm{G}}{\partial A_\mathrm{P}} = -mg\left(\delta_0 A_\mathrm{G} + \zeta_0 \delta_0 \delta_\mathrm{S} A_\mathrm{T}\right)\frac{\partial \rho_\mathrm{P}}{\partial A_\mathrm{P}}$ only if the system were at equilibrium, i.e., if $\delta_\mathrm{S} = e^{-\beta m g \zeta_1 \hat{z}}$.

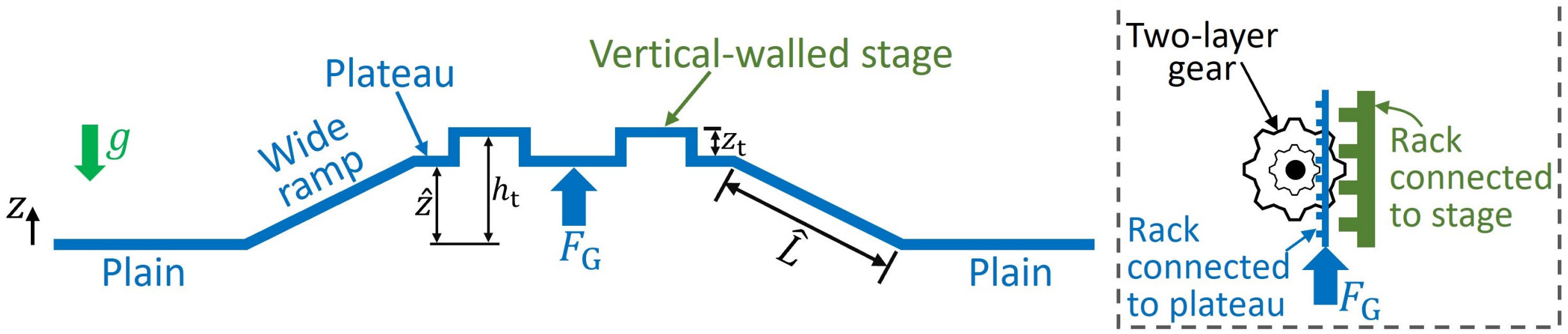

**Fig. 7** A variant system of Figure 6(a). The plateau is connected to the plain through a wide ramp ($\hat{L} \gg \lambda_\mathrm{F}$). A large number of elastic particles move in the gravitation field ($g$). There are vertical-walled stages distributed on the plateau ($z_\mathrm{t} \ll \lambda_\mathrm{F}$). The plateau height ($\hat{z}$) and the stage height ($h_\mathrm{t}$) are adjusted together by the support force ($F_\mathrm{G}$). The inset on the right depicts a mechanism that keeps $z_\mathrm{t} \propto \hat{z}$, using a two-layer gear for each stage.

When $z_\mathrm{t} \ll \lambda_\mathrm{F}$, the stage walls become the SND. As the particle trajectories in the stage walls are locally nonchaotic, $\delta_\mathrm{S} < e^{-\beta m g \zeta_1 \hat{z}}$. Consequently, $\frac{\partial P}{\partial \hat{z}} \neq \frac{1}{\beta}\frac{\partial \rho_\mathrm{P}}{\partial \hat{z}}$, so that in a 4-step isothermal operation cycle of $\hat{z}$ and $A_\mathrm{P}$ similar to Figure 6(c,d), $W_\mathrm{P} \neq W_\mathrm{G}$. If $\delta_\mathrm{S} \approx 1 - \mathrm{erf}\left(\sqrt{\beta m g \zeta_1 \hat{z}}\right)$ (similar to Equation 1), we can derive that

$$-\frac{\partial F_\mathrm{G}}{\partial A_\mathrm{P}} = \Omega_0\left(A_\mathrm{G} + A_\mathrm{T}\delta_2\zeta_0\right) \quad (9)$$

$$\frac{\partial P}{\partial \hat{z}} = \Omega_0\left(A_\mathrm{G} + A_\mathrm{T}\delta_2\xi_\mathrm{P}\right)$$

where $\Omega_0 = Nmg\delta_0(A_\mathrm{P} + \delta_0 A_\mathrm{G} + \delta_0\delta_2 A_\mathrm{T})^{-2}$ and $\xi_\mathrm{P} = 1 - \zeta_1 e^{-\beta m g \zeta_1 \hat{z}}\delta_2^{-1}/\sqrt{\pi\beta m g \zeta_1 \hat{z}}$. Clearly, Equation (7) is not satisfied, i.e., $-\frac{\partial F_\mathrm{G}}{\partial A_\mathrm{P}} \neq \frac{\partial P}{\partial \hat{z}}$, contradicting the second law of thermodynamics. When the change of plateau height ($\mathrm{d}\hat{z} \triangleq \hat{z}_\mathrm{u} - \hat{z}_\mathrm{L}$) and the change of plain area ($\mathrm{d}A_\mathrm{P} \triangleq A_\mathrm{Pu} - A_\mathrm{Pl}$) are small, the total produced work is $W_\mathrm{P} = \frac{\partial P}{\partial \hat{z}}\mathrm{d}\hat{z}\mathrm{d}A_\mathrm{P}$, greater than the total consumed work $W_\mathrm{G} = \left|\frac{\partial F_\mathrm{G}}{\partial A_\mathrm{P}}\right| \mathrm{d}A_\mathrm{P}\mathrm{d}\hat{z}$.

## 6. Concluding Remarks

In essence, what we investigate in this manuscript is an "unusual" particle movement between different energy levels. Across the low-height vertical step, the probability of particle transmission from the low-energy state (the plain) to the high-energy state (the plateau) is non-Boltzmannian. It is not compatible with the entropy statement and the heat-engine statement of the second law of thermodynamics.

In the ideal-gas model in Figure 3(a), a narrow energy barrier is employed as the spontaneously nonequilibrium dimension (SND). The barrier width is much less than the nominal particle mean free path. Inside the SND, the particle-particle collisions are negligible, and the particle trajectories tend to be locally nonchaotic. Everywhere else, the particle movement is ergodic and chaotic. As the nonchaoticity effect "spreads" from the SND to the entire system, a global flow is spontaneously generated from the unforced thermal movement of particles. It leads to entropy decrease without an energetic penalty, which allows for production of useful work in a cycle by absorbing heat from a single thermal reservoir without any other effects. The system contains many particles and can be arbitrarily large. The deviation from thermodynamic equilibrium is steady and significant.

Maxwell's demon is not a SND, since it is not spontaneous. Feynman's ratchet is not a SND, since it is not nonequilibrium. To form a SND in the plateau-plain setup, the nonchaotic process should be local. The globally nonchaotic system of "ghost" particles does not contradict the second law of thermodynamics.

To obtain an appreciable nonequilibrium flow, if the particles are air molecules, at room temperature, the required gravitational acceleration is more than 10 orders of magnitude higher than on Earth. For the experimental study on SND in laboratory, other thermodynamic forces and/or mechanisms need to be explored.

It is worth noting that, although counterintuitive, the intrinsically nonequilibrium steady state is consistent with the basic principle of maximum entropy. It does not violate the fundamental logic that a more probable system state has a higher probability to occur (measured by entropy). The "deterministic" particle movement in the SND imposes additional constraints on the system microstates, i.e., the system becomes less random. Entropy is still maximized, but it reaches a more constrained nonequilibrium maximum ($S_{\text{ne}}$), lower than the less constrained equilibrium maximum ($S_{\text{eq}}$). The classical statements of the second law of thermodynamics may have a boundary. As the theory is expanded to beyond the conventional boundary where the H-theorem does not apply, it could inspire new developments in energy science and technology, new discoveries and explanations of natural phenomena, among others.

## Appendix

### A1. Vertical plane: the non-Boltzmann particle flux ratio

For the vertical plane in Figure 1(a), the analysis in Section 2 is focused on the particle flux ratio ($\delta$). It collectively describes the effects of the particle number density distribution and the particle velocity distribution, directly relevant to the study on the particle flow in Section 3.

Use $\rho_\mathrm{a}(v_\mathrm{z}, z)$ to denote the probability density of finding a particle at height $z$ with the z-component of velocity $v_\mathrm{z}$. In line with Liouville's theorem, since $\frac{\mathrm{d}z}{\mathrm{d}t} = v_\mathrm{z}$ and $\frac{\mathrm{d}v_\mathrm{z}}{\mathrm{d}t} = -g$, $\frac{\partial \rho_\mathrm{a}}{\partial t} = g\frac{\partial \rho_\mathrm{a}}{\partial v_\mathrm{z}} - v_\mathrm{z}\frac{\partial \rho_\mathrm{a}}{\partial z}$ [38]. At the steady state, because $\frac{\partial \rho_\mathrm{a}}{\partial t} = 0$,

$$g\frac{\partial \rho_\mathrm{a}}{\partial v_\mathrm{z}} - v_\mathrm{z}\frac{\partial \rho_\mathrm{a}}{\partial z} = 0.$$

Its general solution is $\rho_\mathrm{a} = f_\mathrm{p}(2gz + v_\mathrm{z}^2)$, where $f_\mathrm{p}$ is a differentiable function.

On the one hand, if the system were at thermodynamic equilibrium, $\rho_\mathrm{a}$ could be written as $\rho_\mathrm{v}(v_\mathrm{z}) \cdot \rho_\mathrm{d}(z)$, where $\rho_\mathrm{v}$ is a function of $v_\mathrm{z}$ and $\rho_\mathrm{d}$ is a function of $z$. The above equation of $\rho_\mathrm{a}$ becomes $\frac{1}{v_\mathrm{z}\rho_\mathrm{v}}\frac{\mathrm{d}\rho_\mathrm{v}}{\mathrm{d}v_\mathrm{z}} = \frac{1}{g\rho_\mathrm{d}}\frac{\mathrm{d}\rho_\mathrm{d}}{\mathrm{d}z} = \hat{C}$, with $\hat{C}$ being a constant. The lower and upper borders of the vertical plane are thermal walls at temperature $T$, and the boundary condition is $\rho_\mathrm{v} \propto e^{-\beta m v_\mathrm{z}^2/2}$ ($z = 0$ or $\hat{z}$). Hence, $\hat{C} = -\beta m$, so that $\rho_\mathrm{v} \propto e^{-\beta m v_\mathrm{z}^2/2}$ and $\rho_\mathrm{d} \propto e^{-\beta m g z}$, and $\rho_\mathrm{a} \propto e^{-\beta m(gz + v_\mathrm{z}^2/2)}$. That is, the system follows the Maxwell-Boltzmann distribution, as expected.

On the other hand, when $\hat{z} \ll \lambda_\mathrm{F}$, $\rho_\mathrm{a}$ may not be expressed as $\rho_\mathrm{v}\rho_\mathrm{d}$, because $v_\mathrm{z}$ and $z$ are not uncorrelated with each other. Furthermore, in addition to the factor of $e^{-\beta m v_\mathrm{z}^2/2}$, the boundary condition at the lower and upper borders should also take into consideration $\rho_\mathrm{a} \propto \bar{v}_\mathrm{z}^{-1}$, with $\bar{v}_\mathrm{z}$ being the average $v_\mathrm{z}$. It reflects that in an area of fast-moving particles, the local particle number density tends to be low, and vice versa. Under this condition, $f_\mathrm{p}$ is not in the form of the Boltzmann factor. As a result, $\delta$ is non-Boltzmannian, which is confirmed by the numerical simulations in Section 2.2 in the main text and Section A4 below.

Another perspective to understand the non-Boltzmann particle flux ratio is as follows. Consider a nonchaotic setup ($\hat{z} \ll \lambda_\mathrm{F}$) wherein the particle-particle collisions are negligible. For an approximate assessment, the particle number density distribution can be calculated as $\bar{\rho}_\mathrm{z} =$ $\int_{\sqrt{2gz}}^{\infty} \kappa_{\mathrm{v}} p_{\mathrm{z}}(v_{\mathrm{z}}) \mathrm{d}v_{\mathrm{z}} = e^{-\beta mgz}$, where $\kappa_{\mathrm{v}} \triangleq v_{\mathrm{z}}/\sqrt{v_{\mathrm{z}}^2 - 2gz}$ represents the aforementioned effect of $\rho_{\mathrm{a}} \propto \bar{v}_{\mathrm{z}}^{-1}$. While $\bar{\rho}_{\mathrm{z}}$ seems in agreement with the Boltzmann factor, the z-component of particle velocity ($v_{\mathrm{z}}$) is nonuniform along $z$. Therefore, the average number of the particles moving across a given horizontal line must be non-Boltzmannian, so is the particle flux ratio ($\delta$). More discussion on the $v - z$ relationship is given in Section 4.3 in the main text.

There are a number of complicated factors in the model systems in Sections 2-5. In the plain and the plateau, near the borders to the step, the particle motion may be locally anisotropic and heterogeneous. Inside the vertical step, the average $v_{\mathrm{z}}$ varies along $z$, while the horizontal component of particle velocity ($v_{\mathrm{y}}$) tends to remain unchanged. If the plateau size ($L_{\mathrm{G}}$) is on the same scale as the particle mean free path, the local particle behavior could exhibit the characteristics of "ballistic transport", affecting the support force ($F_{\mathrm{G}}$). In the current study, we rely on the numerical simulations to quantitatively analyze the system performance. For Figure 3(c) and in Section 5, as a first-order approximation, to obtain analytical solutions, we may assume that the nonequilibrium particle flux ratio ($\delta_*$) is close to $\delta_1$.

A2. Vertical plane: the thermal-wall boundary condition

*A2.i Smooth particle transmission between the vertical section and the horizontal section*

Figure 8(a) depicts a setup of three-dimensional (3D) particle movement that can be analyzed as a two-dimensional (2D) problem. Elastic particles are confined and randomly move in the gap between two frictionless surfaces. The gap thickness ($\hat{t}$) is equal to the particle diameter ($d$). The gap changes direction in the z dimension. The vertical section (the step) is connected to the horizontal section (the plateau and the plain) through curved edges. The radius of curvature of the edges is larger than the particle radius. Thus, the particle movement in the x direction in the horizontal plain/plateau can be smoothly converted to the movement in the z direction in the step, and vice versa. Effectively, the great circles of the particles in parallel with the gap surfaces may be viewed as the hard disks (i.e., the 2D particles) investigated in the main text.

Figure 8(a) is the idealized scenario of Figure 8(b), where the particle size is smaller than the gap thickness ($\hat{t}$). When $\hat{t}$ is much less than the step height ($\hat{z}$), the energy barrier within $\hat{t}$ is negligible (i.e., $mg\hat{t} \ll mg\hat{z}$). The inner surfaces of the gap may be either diffusive or specular. The size of the transition zone ($\bar{x}_{\mathrm{T}}$) is much less than the plateau/plain size but significantly larger than $\bar{y}_{\mathrm{m}} \triangleq 2\sqrt{2/(\beta mg\hat{z})}\,\hat{z}$ (see the discussion on the condition of nonchaoticity in Section 2.1 in the main text), so that the particle behavior in the gravitational field in the transition zone is nearly unrelated to the lateral boundary condition.

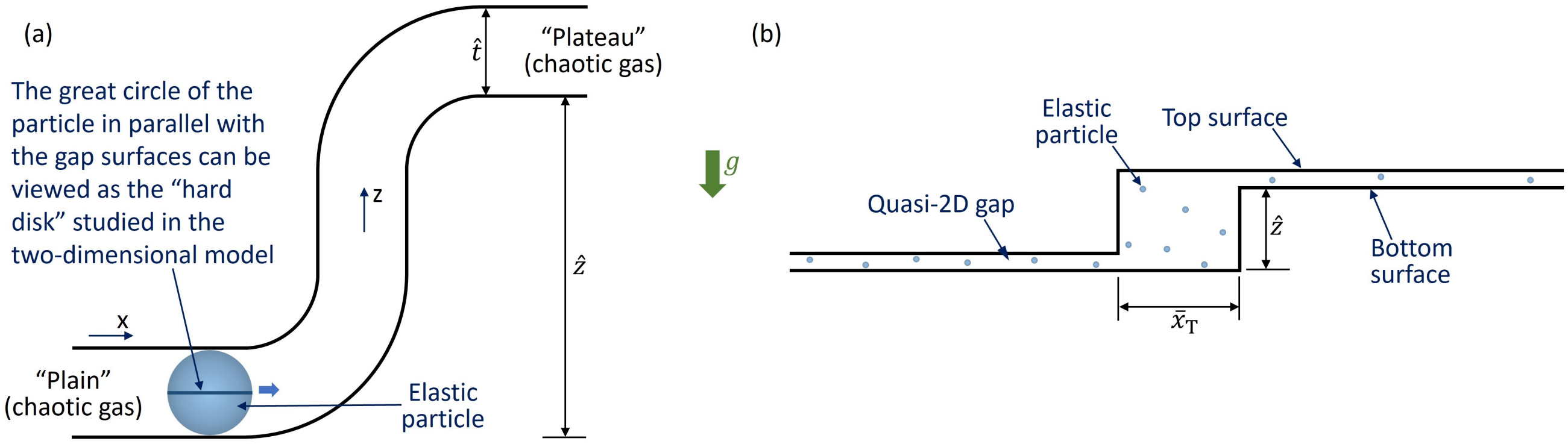

**Fig. 8** Side view of the smooth particle transmission between the horizontal section and the vertical section: **(a)** A setup of 3D particle movement that can be analyzed as a 2D problem. The particle diameter ($d$) is equal to the gap thickness ($\hat{t}$). It is the idealized scenario of **(b)** a less constrained 3D structure, where $d < \hat{t} \ll \hat{z}$ and $\bar{x}_{\mathrm{T}} > \bar{y}_{\mathrm{m}}$.

*A2.ii The thermal-wall boundary condition approximately represents a chaotic gas*

The vertical plane in Figure 1(a) in the main text is a simplified analog to the vertical steps in Figures 3(a), 6(a), and 7. Approximately, in Figure 1(a), the boundary condition of the upper and lower thermal walls represents the effect of the chaotic gas in the large horizontal areas. Both of the thermal walls and the chaotic gas are treated as "black boxes", with the detailed information of microstates being unknown.

In a chaotic gas (e.g., the horizontal plain and plateau in Figure 3a), the particle trajectories are random. No particle can reach the boundary without being interrupted by other particles. For a 2D ideal gas, the expected value of the particle influx at the boundary (i.e., the average number of the particles reaching the boundary) can be written as $\bar{N}_{\mathrm{B}} = \rho_0 L_0 (\bar{v}_{\mathrm{x}} \bar{t}_0)/2$, where $\rho_0$ is the particle number density, $L_0$ is the boundary width, $\bar{t}_0$ is the time duration, $\bar{v}_{\mathrm{x}} = \sqrt{2k_{\mathrm{B}}T/(\pi m)}$ is the average $v_{\mathrm{x}}$, $v_{\mathrm{x}}$ indicates the component of particle velocity normal to the boundary, $k_{\mathrm{B}}$ is the Boltzmann constant, $T$ is temperature, and $m$ is the particle mass. The particle speed ($v$) follows the 2D Maxwell-Boltzmann distribution function $p(v) = \beta m v \cdot e^{-\beta m v^2/2}$ $(v > 0)$, where $\beta = 1/(k_\mathrm{B}T)$ ; $|v_\mathrm{x}|$ follows the one-dimensional Maxwell-Boltzmann distribution $p_\mathrm{x}(v_\mathrm{x}) = \sqrt{2\beta m/\pi}\, e^{-\beta m v_\mathrm{z}^2/2}$. The gas pressure on the boundary (i.e., the ideal-gas law) can be derived as

$$P = \frac{\bar{N}_\mathrm{B} \cdot \int_0^\infty \int_0^{\pi/2} (2mv \cdot \cos\theta) \cdot p(v) \mathrm{d}\theta \mathrm{d}v}{L_0 \bar{t}_0} = \rho_0 k_\mathrm{B} T,$$

where $\theta$ is the incident angle. If a temperature sensor is placed at the gas-container wall, the measured temperature is

$$\bar{T}_\mathrm{wall} = \frac{1}{k_\mathrm{B}} \left[ \frac{\bar{N}_\mathrm{B} \cdot \int_0^\infty \frac{mv^2}{2} p(v) \mathrm{d}v}{\bar{N}_\mathrm{B}} \right] = T,$$

as it should be. It can be seen that in terms of the particle flux across the plain/plateau-step border, the chaotic-gas boundary condition of $p(v)$ and a thermal wall are equivalent to each other.

In Figure 3(a), if the gas phase in the horizontal areas were nonchaotic, its boundary condition with the step may not be approximated by a thermal wall. Without particle-particle collision, the particle flux toward the boundary should not be analyzed by the mean-field theory using $\bar{v}_\mathrm{x}$, but by accounting for individual trajectories: $\bar{N}_\mathrm{B}' = \rho_0 L_0 \bar{t}_0 \cdot \frac{1}{2} \int_0^\infty [v_\mathrm{x} \cdot p_\mathrm{x}(v_\mathrm{x})] \mathrm{d}v_\mathrm{x}$. While it gives an ideal-gas-like equation of pressure

$$P = \frac{\rho_0 L_0 \bar{t}_0 \frac{1}{2} \int_0^\infty (2mv_\mathrm{x}) v_\mathrm{x} p_\mathrm{x} \mathrm{d}v_\mathrm{x}}{L_0 \bar{t}_0} = \rho_0 k_\mathrm{B} T,$$

when calculating the effective temperature of the particles reaching the boundary, it leads to

$$\bar{T}_\mathrm{wall} = \frac{1}{k_\mathrm{B}} \left[ \frac{\rho_0 L_0 \bar{t}_0 \cdot \frac{1}{2} \int_0^\infty \frac{m v_\mathrm{x}^2}{2} v_\mathrm{x} p_\mathrm{x}(v_\mathrm{x}) \mathrm{d}v_\mathrm{x}}{\bar{N}_\mathrm{B}'} + \frac{k_\mathrm{B} T}{2} \right] = \frac{3}{2} T, \text{ or equivalently,}$$

$$\bar{T}_\mathrm{wall} = \frac{1}{k_\mathrm{B}} \left[ \frac{\rho_0 L_0 \bar{t}_0 \cdot \frac{1}{2} \int_0^\infty \int_0^{\pi/2} \left( \frac{mv^2}{2} \right) [v \cdot \cos\theta \cdot p(v)] \mathrm{d}\theta \mathrm{d}v}{\rho_0 L_0 \bar{t}_0 \cdot \frac{1}{2} \int_0^\infty \int_0^{\pi/2} [v \cdot \cos\theta \cdot p(v)] \mathrm{d}\theta \mathrm{d}v} \right] = \frac{3}{2} T,$$

where the term of $k_\mathrm{B}T/2$ in the first equation is the average particle kinetic energy in the dimension parallel to the boundary; the denominator in the bracket in the second equation is $\bar{N}_\mathrm{B}'$. Such a thermally nonequilibrium $\bar{T}_\mathrm{wall}$ has been well known for nonchaotic gases [e.g., 21,22], but is irrelevant to a chaotic gas. Moreover, we can verify that

$$\frac{1}{2} \int_{\sqrt{2g\hat{z}}}^\infty [v_\mathrm{x} \cdot p_\mathrm{x}(v_\mathrm{x})] \mathrm{d}v_\mathrm{x} = e^{-\beta m g \hat{z}} \cdot \frac{1}{2} \int_0^\infty [v_\mathrm{x} \cdot p_\mathrm{x}(v_\mathrm{x})] \mathrm{d}v_\mathrm{x},$$

where $mg\hat{z}$ is the gravitational energy barrier. It suggests that if the step-ramp model system in Figure 3(a) is not locally but globally nonchaotic (i.e., particle-particle collision is negligible not only in the vertical step, but also in the plain, the plateau, and the ramp), the particle distribution tends to follow the Boltzmann factor, which is consistent with the reference numerical tests on "ghost" particles (the red data points in Figure 4b in the main text and Figure 11a below).

## A3. Introduction to the algorithm of computer simulation

The particles are 2D hard disks. In between collisions, the particle trajectories are governed by $\mathrm{d}\vec{v} = \vec{a}\Delta t_0$, where $\mathrm{d}\vec{v}$ is the change in particle velocity, $\vec{a}$ is the acceleration, and $\Delta t_0$ is the timestep. On the plain and the plateau, $\vec{a} = 0$. In the vertical plane/step, $|\vec{a}| = g$. In the wide ramp, $|\vec{a}| = \hat{z}g/\hat{L}$.

The particle collisions are elastic, calculated by solving Newton's equations (conservation of energy and momentum): $v_1^2 + v_2^2 = v_1^{*2} + v_2^{*2}$, $v_{\mathrm{n}1} + v_{\mathrm{n}2} = v_{\mathrm{n}1}^* + v_{\mathrm{n}2}^*$, and $v_{\mathrm{t}1} = v_{\mathrm{t}1}^*$ and $v_{\mathrm{t}2} = v_{\mathrm{t}2}^*$, where subscripts "1" and "2" indicate two particles, subscripts "n" and "t" respectively indicate the collision (center-to-center) direction and the tangential direction, and the asterisk indicates "after collision" (absence of it indicates "before collision").

Particle collisions happen in the middle of the timesteps. We use a high time resolution so that in each timestep, the expected value of the particle displacement is less than 5% of the particle size. At the onset of a timestep, the program first predicts the virtual position of every particle at the end of the timestep, as if there were no particle collision. Collision is defined as the particle-particle or particle-wall overlap. If no collision takes place, the simulation will continue to the next timestep. Otherwise, the exact time of the collision will be identified, and the particle velocities and locations will be calculated and updated.

The Matlab code contains two programs [17]. The first program (dragons_egg.m) defines the basic settings, such as $d$, $m$, $\Delta t_0$, and the shape and size of the simulation box. The second program (ball_collision.m) is a function that receives the parameters from the first program and runs the system stepwise. It uses the check_overlap function to determine the collision time, locate the involved particles, and update their information.

## A4. Vertical plane: various initial and boundary conditions

We tested various boundary conditions and initial conditions for the system in Figure 1(a). As long as the particle-particle collisions are negligible in the vertical plane, the steady-state particle flux ratio is always non-Boltzmannian.

For $\hat{z}/\lambda_{\mathrm{F}} \approx 0.1$, two different initial conditions are investigated. The first is near-uniform (Figure 9a), the same as in Section 2.2; i.e., initially, the particles are randomly placed in the plane. The second is near-equilibrium (Figure 9b); i.e., the initial probability for a particle to be placed at height $z$ is proportional to the Boltzmann factor, $e^{-\beta mgz}$. The initial particle speed randomly follows $p(v)$; the initial particle direction is random.

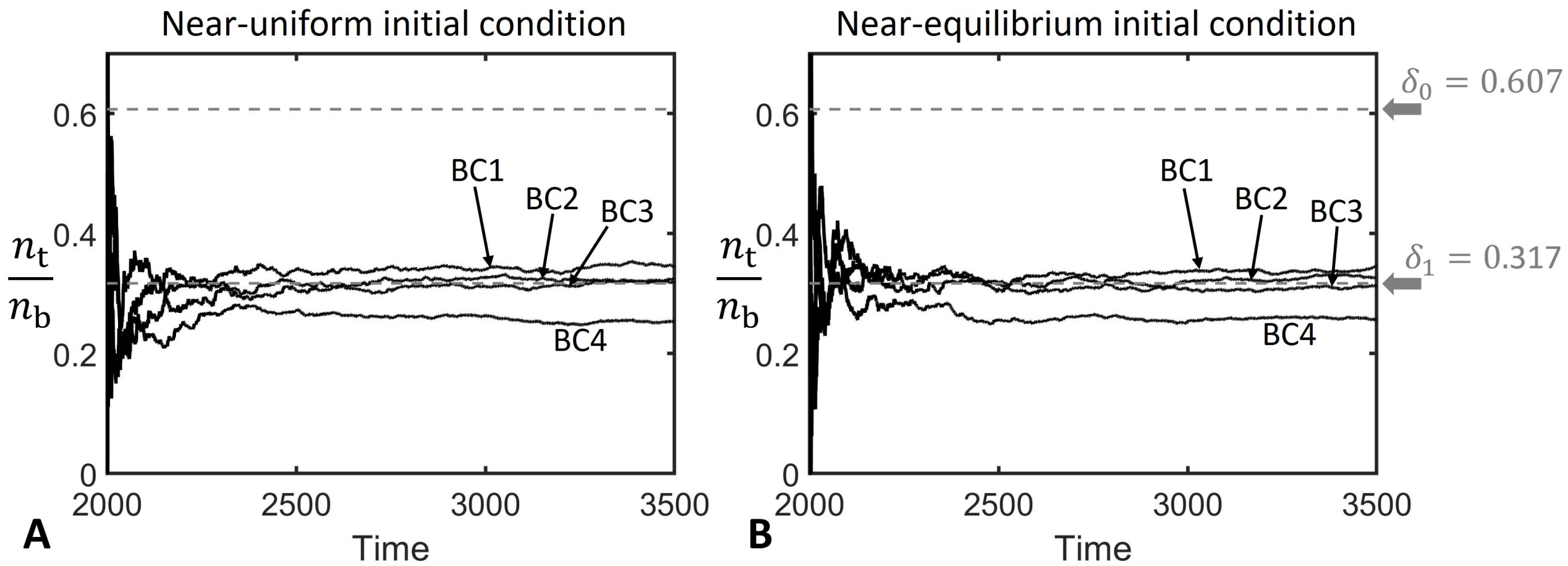

**Fig. 9** For the system in Figure 1(a), two different initial conditions are tested ($\hat{z}/\lambda_{\mathrm{F}} = 0.1$): initially, the particles are randomly placed in the vertical plane, with the probability density at height $z$ **(A)** being uniform, or **(B)** following the Boltzmann factor, $e^{-\beta mgz}$. For each initial condition, four different boundary conditions are tested: BC1, BC2, BC3, and BC4. In all the simulation cases, the steady-state $n_{\mathrm{t}}/n_{\mathrm{b}}$ ratio is much less than the Boltzmann factor ($\delta_0$).

The lateral borders (DC and D′C′) are open and use periodic boundary condition. For each initial condition, we investigate four different boundary conditions at the upper/lower borders (DD′ and CC′): BC1, BC2, BC3, and BC4. BC1 is the same as the boundary condition used in Section 2.2. Both of the upper and lower borders are thermal walls at the same temperature; the reflected particle direction is random; the speed of the reflected particles is not correlated with the incident speed, but randomly follows $p(v)$. BC2 and BC3 have the same bottom boundary condition as that of BC1. The top boundary of BC2 is a diffusive wall, with the reflected particle speed being the same as the incident speed. The top boundary of BC3 is a specular wall. In BC4, both of the upper and lower boundaries are the same diffusive walls as the upper border of BC2; i.e., the reflected particle direction is random, and the reflected particle speed is equal to the incident speed. BC4 represents an isolated setup.

All the other parameters and procedures are the same as in Section 2.2. Figure 9 shows typical time profiles of the running average of $n_t/n_b$; the steady-state $n_t/n_b$ indicates the particle flux ratio ($\delta$). It can be seen that for all the boundary conditions and both initial conditions, $\delta$ is significantly smaller than the Boltzmann factor, $\delta_0$. For BC1, BC2, and BC3, $\delta$ is close to or slightly larger than $\delta_1$. For BC4, $\delta$ is lower than $\delta_1$, which may be associated with the heterogenous particle velocity distribution along $z$. For each boundary condition, the initial conditions have little influence on the steady-state $n_t/n_b$.

A5. Inspiration from Feynman's ratchet

As depicted in Figure 10, Feynman's ratchet is two-ended [7]. One end is a set of vanes, and the other end is a set of ratchet and pawl. They are connected through a rigid rod. Due to the random impact of surrounding gas molecules, the vanes undergo a rotational Brownian movement. At first glance, it seems that the ratchet might selectively guide the oscillation steps, so that the vanes are only allowed to rotate in the forward direction. Yet, such a "perpetual motion machine" would not work. To overcome the energy barrier of the spring of the pawl ($\Delta E_p$), the probabilities for both of the vanes and the ratchet are governed by the same Boltzmann factor, $e^{-\beta\cdot\Delta E_p}$. Thus, the overall motions of the ratchet and the vanes counterbalance each other. Mere geometric asymmetry does not cause any anomalous effect.

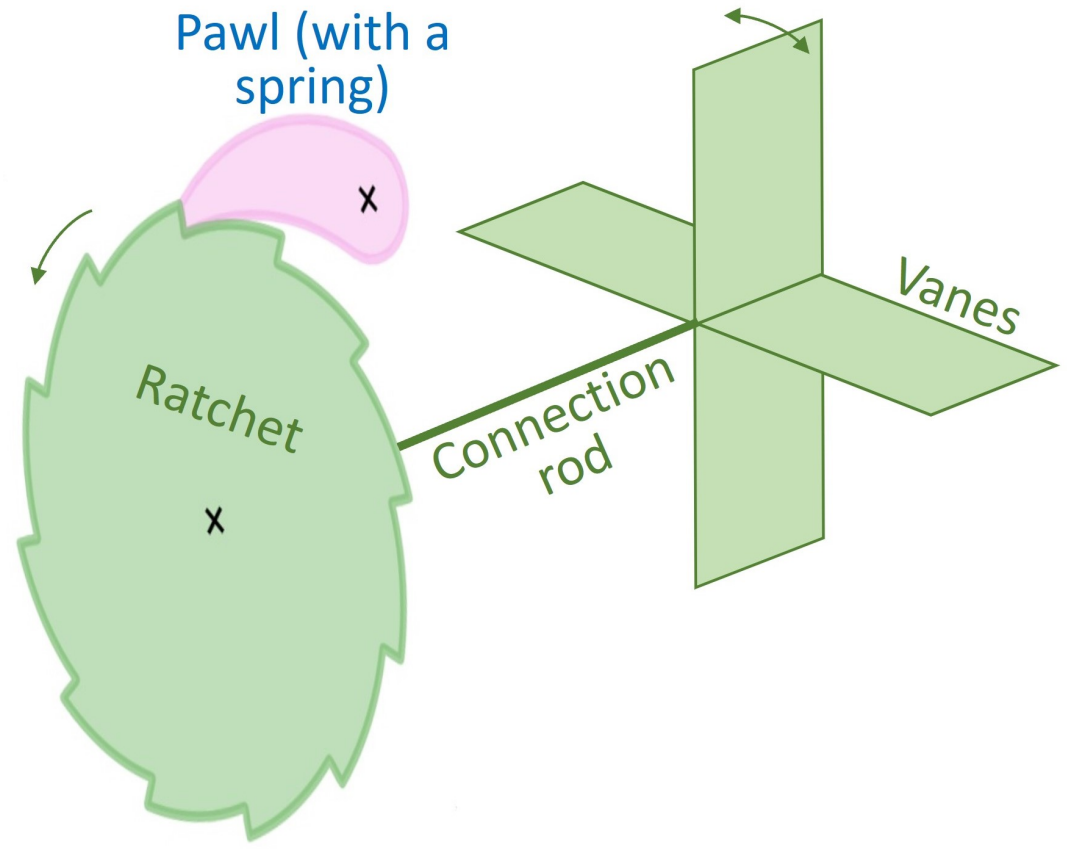

**Figure 10** Schematic of Feynman's ratchet. The spring of the pawl is not shown.

As analyzed in Section 2 in the main text, without extensive particle-particle collision, the steady state of a nonchaotic vertical plane may be intrinsically nonequilibrium, which raises a critical question: In a two-ended system, what would happen if one end tends to reach thermodynamic equilibrium, while the other end does not? Such a structure could be unbalanced.

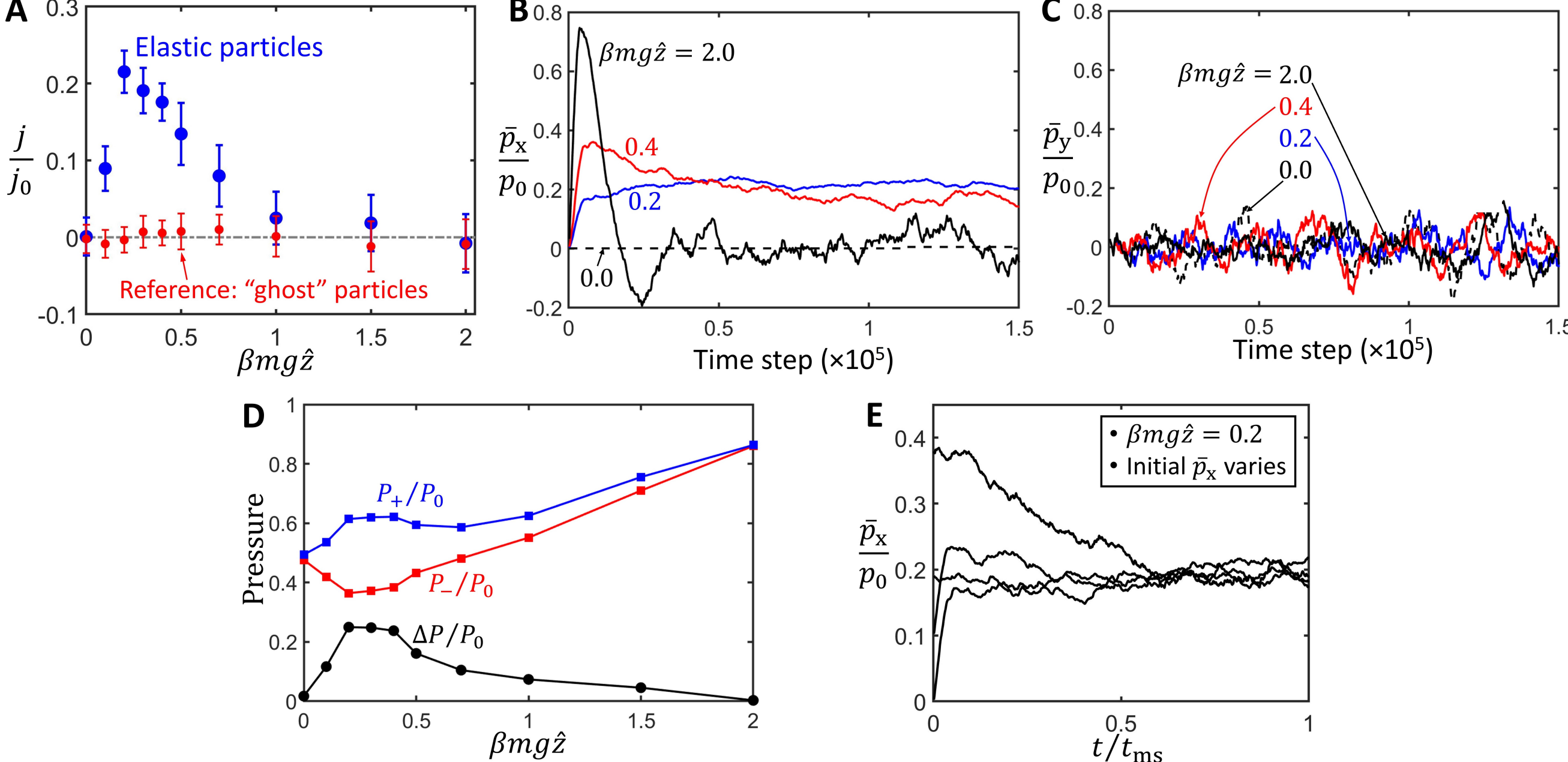

**Fig. 11 (a)** MC simulation results for the specular-wall boundary condition and the on-plain initial condition: the calculated steady-state particle flow rate ($j$) as a function of $\beta mg\hat{z}$. The red data points show the reference tests on "ghost" particles, with the particle-particle collision being turned off. **(b)** Typical time profiles of the average x-component of particle momentum ($\bar{p}_x$) and **(c)** the average y-component of particle momentum ($\bar{p}_y$). **(d)** The inner pressure at the left/right border (AA′ and BB′). **(e)** Typical time profiles of $\bar{p}_x$, with various initial values. The system setting is the same as in Section 3.2 (except for the initial $\bar{p}_x$). The steady-state $\bar{p}_x$ always converges to the same level. Time ($t$) is normalized by the total simulation time ($t_{ms}$). When the initial $\bar{p}_x/p_0 \approx 0.4$, $t_{ms} = 3\times10^5$; for all the other curves, $t_{ms} = 1.5\times10^5$.

<u>A6. Step-ramp system: various initial and boundary conditions</u>

For the step-ramp model system in Figure 3(a), we tested different initial and boundary conditions. Figure 11(a-d) shows the MC simulation result when the upper/lower borders (AB and A′B′) are rigid specular walls; the left/right borders are open and use periodic boundary condition. Initially, the particles are randomly placed on the plain; the step, the ramp, and the plateau are empty. The red data points in Figure 11(a) are for the reference tests on "ghost" particles, with the particle-particle collision being turned off; the particles can be reflected by the specular walls. All the other settings are the same as in Section 3.2. It can be seen that the main characteristics of Figure 11(a-d) are similar to those of Figure 4 (b-e), suggesting that the nonequilibrium nature of the steady state is insensitive to the specific forms of boundary/initial condition under investigation.

In a numerical experiment of $\beta mg\hat{z} = 0.2$, we use the same initial condition as in Figure 11(a-d) (i.e., initially, the particles are randomly placed on the plain), with everything else being the same as in Section 3.2, including the boundary condition (i.e., all the borders use periodic boundary condition). The calculated steady-state particle flux is $j/j_0 = 0.2011 \pm 0.0400$, close to the results in Figure 4(b) and Figure 11(a); the data range indicates the 90%-confidence interval.

In another set of numerical experiments of $\beta mg\hat{z} = 0.2$, we examine the effect of the initial particle flow rate, as shown in Figure 11(e). A number of configurations are randomly generated similarly to Figure 4(c). Different biases are selected, so that the initial $\bar{p}_\mathrm{x}/p_0$ varies in the range from 0 to 0.4; everything else is the same as in Figure 4(c). The initial $\bar{p}_\mathrm{y}$ is near zero. The total system energy ($E_\mathrm{tot}$) is close to $Nk_\mathrm{B}T$, with the difference less than 0.2%. The results indicate that at the steady state, $\bar{p}_\mathrm{x}$ always converges to the same level.

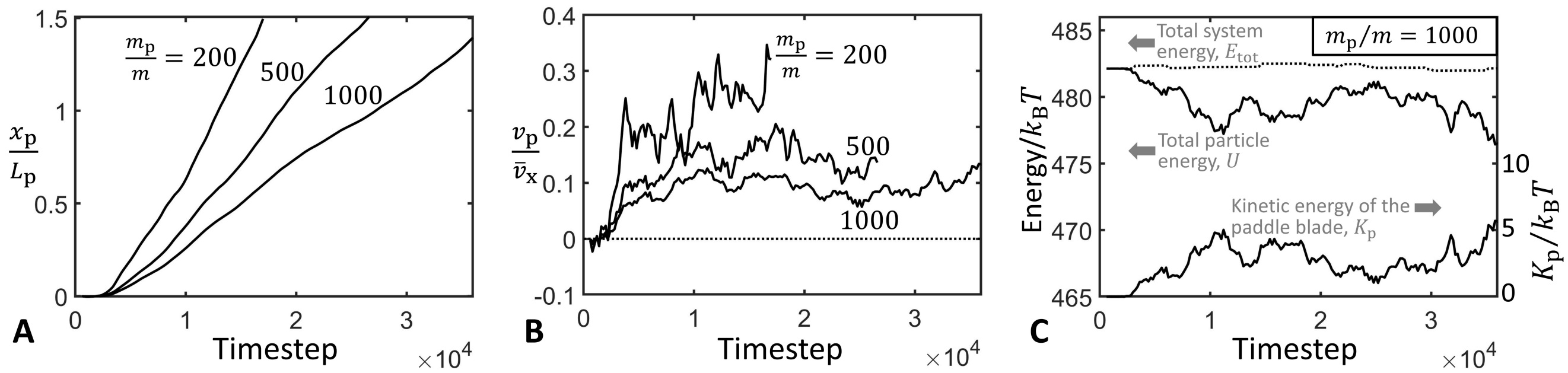

**Fig.12** Simulation results for the specular-wall boundary condition. Typical time profiles of **(a)** the displacement ($x_\mathrm{p}$), **(b)** the velocity ($v_\mathrm{p}$), and **(c)** the kinetic energy ($K_\mathrm{p}$) of the paddle blade.

A7. Paddle blade: various initial and boundary conditions

Figure 12 shows the MC simulation result of the motion of the paddle blade, with a different boundary and initial condition from Figure 5. The upper and lower borders of the simulation box are specular walls; the initial configuration comes from the steady state in Figure 11(a) ($\beta mg\hat{z} = 0.2$). All the other settings and procedure are the same as in Section 3.3. It can be seen that the main characteristics of Figure 12(a-c) are similar to those of Figure 5(b-d), suggesting that the behavior of the paddle blade is not sensitive to the boundary and initial conditions under investigation.

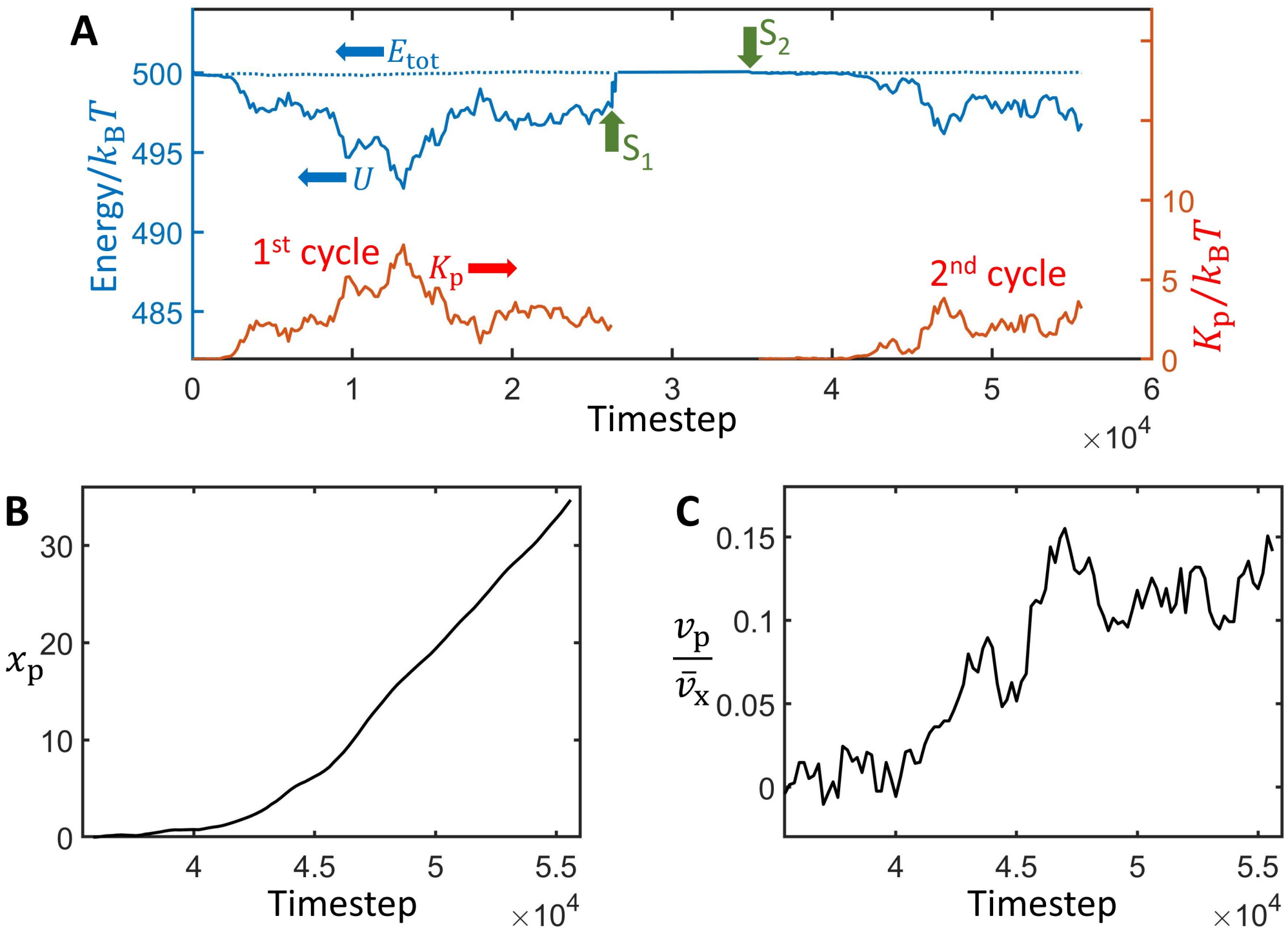

**Fig. 13 (a)** Two cycles of thermal-to-mechanical energy conversion: typical time profiles of the total particle energy ($U$), the total system energy ($E_{tot}$), and the paddle-blade kinetic energy ($K_p$). Arrow "$S_1$" indicates the end of the first cycle and the beginning of heating; arrow "$S_2$" indicates the beginning of the second cycle. The time between "$S_1$" and "$S_2$" is for heating (400 timesteps) and randomization (8800 timesteps). **(b)** Typical time profiles of the displacement ($x_p$) and **(c)** the velocity ($v_p$) of the paddle blade in the second cycle. The data of the first cycle are shown in Figure 5 in the main text ($m_p/m = 500$).

## A8. Paddle blade: the second cycle of thermal-to-mechanical energy conversion

At the end of the simulation in Figure 5(d) (the first cycle of thermal-to-mechanical energy conversion), the paddle blade is removed, and a heat exchanger is placed in the middle of the simulation box. The heat exchanger is modeled as a rigid line perpendicular to the x axis, with the length of $w_0$. When a particle collides with the heat exchanger, it would be reflected to a random direction; the speed of the reflected particle ($v_{re}$) is not correlated with the incident speed, but randomly follows the 2D Maxwell-Boltzmann distribution $p(v_{\mathrm{re}}) = (\beta_{\mathrm{h}} m v_{\mathrm{re}}) e^{-\beta_{\mathrm{h}} m v_{\mathrm{re}}^2/2}$, where $\beta_{\mathrm{h}} = 1/(k_{\mathrm{B}} T_{\mathrm{r}})$, and $T_{\mathrm{r}} = 1200$ is the effective heat-exchanger temperature.

After 400 timesteps, the heat exchanger is removed. We performed 100 random numerical tests on the heating process. The expected value of the total particle energy ($U$) after heating was $500.1 k_{\mathrm{B}} T$ ($T = 1000$), close to the total system energy in Figure 5(d) ($\sim 500 k_B T$). Figure 13(a) shows typical time profiles of energy evolution.

The system is randomized for 8800 timesteps, after which the paddle blade ($m_{\mathrm{p}}/m = 500$) is placed back to the middle of the left plane. The displacement and the velocity of the paddle blade are observed, as shown in Figure 13(b,c). It can be seen that with a single thermal reservoir (the heat exchanger), the thermal-to-mechanical energy conversion can be operated cyclically, without any other effect.

A9. Rearrangement of the step and the ramp

Figure 14 is a variant of Figure 3(a). The system is closed and immersed in a thermal bath. If both of the low-height step ($\hat{z} \ll \lambda_{\mathrm{F}}$) and the wide ramp ($\hat{L} \gg \lambda_{\mathrm{F}}$) are open, there would be a continuous circular particle flow between the upper plateau and the lower plain. Here, we use a frictionless sliding door to alternately open and close the step and the ramp. The sliding door is a macroscopic object. Unlike Feynman's ratchet, its thermal vibration is negligible.

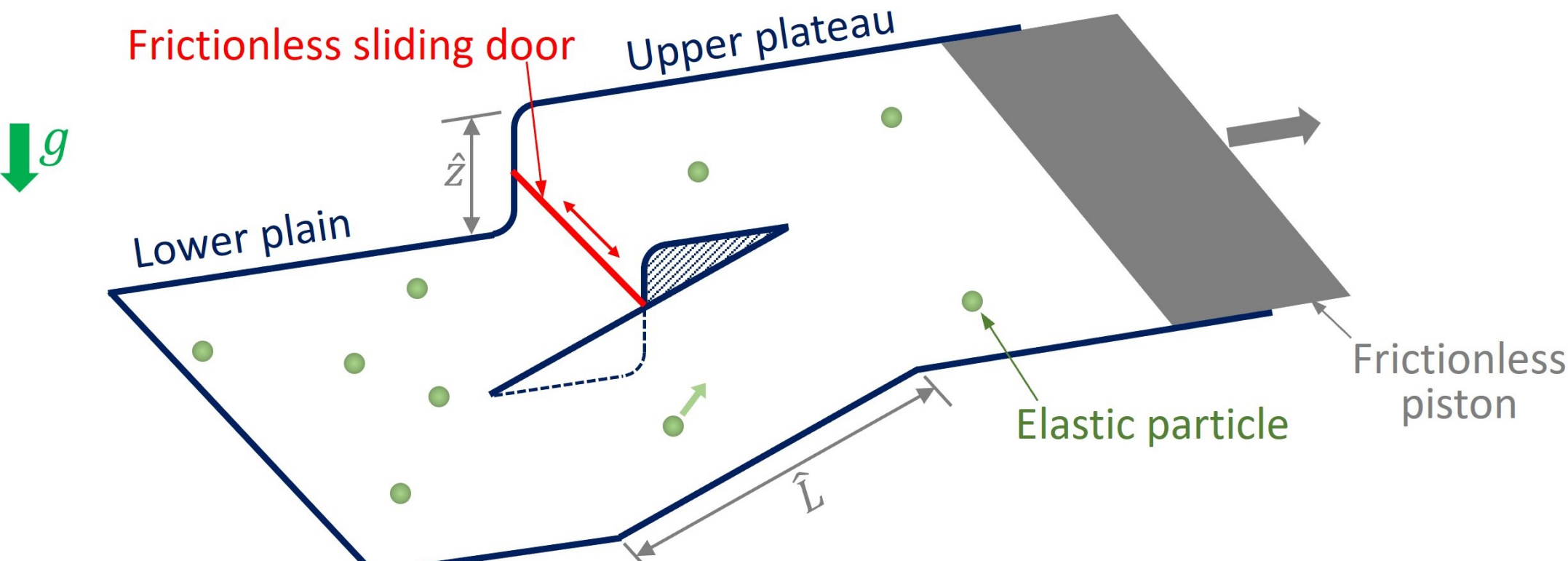

**Fig. 14** A variant system of Figure 3(a). The plateau and the plain are connected partly through a low-height vertical step ($\hat{z} \ll \lambda_{\mathrm{F}}$) and partly through a wide ramp ($\hat{L} \gg \lambda_{\mathrm{F}}$). The step and the ramp are alternately opened and closed by a frictionless sliding door. Correspondingly, a frictionless piston expands and compresses the plateau.

When the ramp is open and the step is closed, the system is at thermodynamic equilibrium. The plateau-to-plain particle number density ratio is $\delta_0$. The plateau area is expanded by a frictionless piston, doing work $W_{\mathrm{P1}}$ to the environment. Then, the ramp is closed and the step is opened, so that the plateau-to-plain particle number density ratio is $\delta_*$. Under this condition, the piston compresses the plateau back to the original size, consuming work $W_{\mathrm{P2}}$. Because $\delta_0 > \delta_*$, $W_{\mathrm{P1}} > W_{\mathrm{P2}}$. The work production can be operated cyclically, by absorbing heat from a single thermal reservoir (the environment).

A10. Considerations on charge carriers

One possible approach of experimental study of SND is to use a mesoscopic physical system. For example, consider a conductive or semiconductive nanowire or nanolayer in an external electric field. It has an upper shelf (the plateau) and a lower shelf (the plain), connected by a nanostep. The nanostep size ($\hat{z}$) is much smaller than the mean free path of the charge carriers. In a regular setup, as the "excess" charge carriers tend to repel each other, the potential along the surface is constant, corresponding to the zero-gravity condition in Figure 3. Thus, the ballistic transport of the charge carriers between the two shelves does not cause any peculiar consequence. To design a highly nonequilibrium setup, it would be interesting to investigate whether a significant potential difference might be maintained across the nanostep. To minimize the shielding effect, a small thickness and a small plateau/stage size may be helpful. Desirably, they should be less than the average spacing of charge carriers or the Debye length. The top and bottom facets may have different features. The thickness can be uneven, so that $\hat{z}$ is not the same at the two sides. Multiple nanosteps can be placed in tandem and/or in parallel. If the "excess" charge carriers fill more low-energy states on one shelf and the reverse process happens on the other shelf, the temperature field might be heterogeneous, resulting in unusual thermal phenomena.

In a metal, if a nonequilibrium electron flow could be spontaneously induced, the upper limit of the power density might be more than 10 kW/cm$^3$. This assessment is based on the observation that the peak $j$ in Figure 4(b) is ~20% of $j_0$, with the assumption that such a ratio is also relevant to the charge carriers. As a qualitative comparison with the billiard-type system, for the sake of simplicity, here we only account for the high-energy charge carriers with the number density ($\rho_e$) approximately following the Maxwell-Boltzmann distribution. Their energy density is on the scale of $E_F \rho_e$, where $E_F$ is the Fermi energy (on the scale of a few eV) and $\rho_e$ is assumed to be $10^{20}$ m$^{-3}$. The characteristic time is taken as $10^6 t_{ce}$, where $t_{ce} = \hat{z}/v_F$, and $v_F$ is the Fermi velocity (on the scale of $10^6$ m/s).

Compared to the particle flow in Figure 3, experimental realization of the concept in Figure 6(a) or Figure 7 is probably easier, e.g., by using an electrolyte solution, a bent graphene layer, or a compound conductor-insulator-semiconductor configuration. Geometry is not the only mechanism that can lead to different energy states. For instance, if the external electric field is nonuniform and/or contains double-layer-like structures, adjusting the local potential difference may be analogous to controlling the plateau/stage height. Moreover, following the fundamental principle of [16], entropy-barrier-type asymmetry might be achieved, e.g., based on the asymmetric inelastic large-angle scattering, possibly in a low-dimensional setting. The key factors and their efficacy remain to be seen.

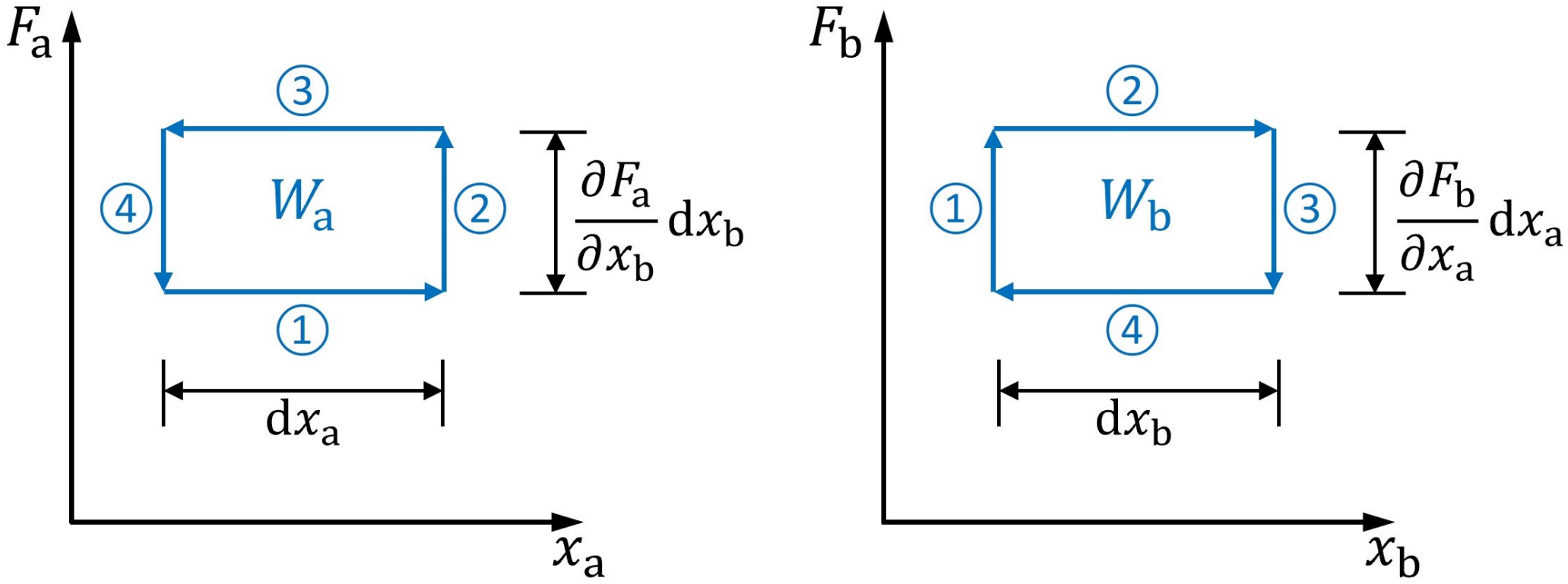

**Fig. 15** A four-step isothermal cycle. The circled numbers indicate the operation steps. Two thermally correlated thermodynamic forces ($F_a$ and $F_b$) are adjusted alternately. $F_a$ consumes work $W_a$; $F_b$ produces work $W_b$. Without loss of generality, here the cross-influence is shown as positive; $dx_a$ and $dx_b$ are arbitrarily small.

A11. Cross-influence of thermally correlated thermodynamics forces

Consider a closed system immersed in a thermal bath. It has two thermally correlated thermodynamic forces, $F_a$ and $F_b$. Their conjugate variables are $x_a$ and $x_b$, respectively. The cross-influence of $F_a$ and $F_b$ is defined as $F'_{ab} \triangleq \frac{\partial F_a}{\partial x_b}$ and $F'_{ba} \triangleq \frac{\partial F_b}{\partial x_a}$. The system is operated in the 4-step isothermal cycle shown in Figure 15. In step 1, $x_a$ is increased by a small amount $dx_a$. It causes $F_{\mathrm{b}}$ to change by $F'_{\mathrm{ba}}\mathrm{d}x_{\mathrm{a}}$, and does work $F_{\mathrm{a}}\mathrm{d}x_{\mathrm{a}}$ to the environment. In step 2, $x_{\mathrm{b}}$ is increased by a small amount $\mathrm{d}x_{\mathrm{b}}$, which causes $F_{\mathrm{a}}$ to vary by $F'_{\mathrm{ab}}\mathrm{d}x_{\mathrm{b}}$, and does work $(F_{\mathrm{b}} + F'_{\mathrm{ba}}\mathrm{d}x_{\mathrm{a}})\mathrm{d}x_{\mathrm{b}}$ to the environment. Then, in step 3, $x_{\mathrm{a}}$ is reduced by $\mathrm{d}x_{\mathrm{a}}$; $F_{\mathrm{b}}$ changes back, and the system consumes work $(F_{\mathrm{a}} + F'_{\mathrm{ab}}\mathrm{d}x_{\mathrm{b}})\mathrm{d}x_{\mathrm{a}}$. Finally, in step 4, $x_{\mathrm{b}}$ is decreased by $\mathrm{d}x_{\mathrm{b}}$; the system consumes work $F_{\mathrm{b}}\mathrm{d}x_{\mathrm{b}}$, and returns to its initial condition.

Overall, $F_{\mathrm{a}}$ consumes work $W_{\mathrm{a}} = F'_{\mathrm{ba}}\mathrm{d}x_{\mathrm{b}}\mathrm{d}x_{\mathrm{a}}$, and $F_{\mathrm{b}}$ produces work $W_{\mathrm{b}} = F'_{\mathrm{ab}}\mathrm{d}x_{\mathrm{a}}\mathrm{d}x_{\mathrm{b}}$. The heat-engine statement of the second law of thermodynamics demands that $W_{\mathrm{a}} = W_{\mathrm{b}}$, which leads to $F'_{\mathrm{ab}} = F'_{\mathrm{ba}}$. It is Equation (7) in the main text.

A12. SND-induced additional constraint

Consider a canonical ensemble of the plateau-plain system of two-dimensional elastic particles in Figure 6(a). The plateau size and the plain size are much larger than the step size. To demonstrate the basic concept, for an approximate assessment of entropy, we assume that the temperature field is uniform, and ignore the low-probability thermally nonequilibrium microstates, similarly to the Lagrange multiplier analysis in [16]. At thermal equilibrium, the internal energy is $U = Nk_{\mathrm{B}}T + N_{\mathrm{G}}mg\hat{z}$, where $N$ is the total particle number, $k_{\mathrm{B}}$ is the Boltzmann constant, $T$ is temperature, $N_{\mathrm{G}}$ is the particle number on the plateau, $m$ is the particle mass, $g$ is the gravitational acceleration, and $\hat{z}$ is the plateau height. The system entropy is

$$S \triangleq -k_{\mathrm{B}} \sum_{\chi} f_{\chi} \ln f_{\chi} \approx -k_{\mathrm{B}} \sum_{n} g_n (f_n \ln f_n),$$

where $f_{\chi}$ is the probability of the $\chi$-th possible microstate, and $f_n$ and $g_n$ are the probability and the density of states at the $n$-th energy level $\epsilon_n$, respectively. The index of the energy level ($n$) may be taken as $N_{\mathrm{G}}$, with $\epsilon_n \approx Nk_{\mathrm{B}}T + nmg\hat{z}$ ($n = 0,1,2 \dots N$).

If $\hat{z} \gg \lambda_{\mathrm{F}}$, the system is chaotic. There are two constraints on $f_n$:

$$\sum_{n=0}^{N} g_n f_n = 1$$

$$\sum_{n=0}^{N} g_n f_n \epsilon_n = \overline{U}_{\mathrm{n}}$$

where $\overline{U}_{\mathrm{n}}$ is the expected value of the internal energy. Define the Lagrangian as

$$\mathcal{L}_0 = -\sum_{n=0}^{N} g_n (f_n \ln f_n) + \lambda_2 \left(1 - \sum_{n=0}^{N} g_n f_n\right) + \bar{\beta}_2 \left(\overline{U}_{\mathrm{n}} - \sum_{n=0}^{N} g_n f_n \epsilon_n\right),$$

with $\lambda_2$ and $\bar{\beta}_2$ being the Lagrange multipliers. The principle of maximum entropy requires that

$$\frac{\partial \mathcal{L}_0}{\partial f_n} = 0,$$

which leads to $f_n \propto e^{-\beta \epsilon_n}$, so that

$$f_n \propto e^{-\beta n m g \hat{z}}$$

where $\beta = 1/(k_\mathrm{B} T)$. As expected, $f_n$ is governed by the Boltzmann factor ($\delta_0 \triangleq e^{-\beta m g \hat{z}}$). It corresponds to the maximum possible equilibrium entropy, $S_\mathrm{eq}$.

If $\hat{z} \ll \lambda_\mathrm{F}$, the system is locally nonchaotic. The "deterministic" particle movement in the low-height vertical step (the SND) imposes additional constraints on the system microstates:

$$\frac{f_{n_1}}{f_{n_2}} = \delta_*^{\Delta n_{12}}$$

for all the possible combinations of $n_1$ and $n_2$, where subscripts "$n_1$" and "$n_2$" indicate two energy levels ($n_1, n_2 = 0,1,2, \dots N$), $\Delta n_{12}$ is the difference in $N_\mathrm{G}$ between them, and $\delta_*$ is the particle flux ratio across the step. It could be understood by considering that, in terms of the distribution of particle number density, the effect of the locally nonchaotic step is equivalent to a chaotic step with the energy barrier of $k_\mathrm{B} T \ln \delta_*^{-1}$. If $\delta_* = \delta_0$, the above constraint is trivial, since $f_{n_1}/f_{n_2} = \delta_0^{\Delta n_{12}}$ can be derived from $f_n \propto e^{-\beta n m g \hat{z}}$, based on the conventional constraints ($\sum_{n=0}^{N} g_n f_n = 1$ and $\sum_{n=0}^{N} g_n f_n \, \epsilon_n = \overline{U}_\mathrm{n}$). As $\delta_* \neq \delta_0$, it provides useful information, and the system microstates become less random. The Lagrangian should be redefined as

$$\mathcal{L}_1 = -\sum_{n=0}^{N} g_n (f_n \ln f_n) + \lambda_3 \left(1 - \sum_{n=0}^{N} g_n f_n \right) + \overline{\beta}_3 \left(\overline{U}_\mathrm{n} - \sum_{n=0}^{N} g_n f_n \, \epsilon_n\right) + \sum_{n_1, n_2 = 0}^{N} \overline{\mu}_{12} \left(f_{n_1} - \delta_*^{\Delta n_{12}} f_{n_2}\right)$$

with $\lambda_3$, $\overline{\beta}_3$, and $\overline{\mu}_{12}$ being the Lagrange multipliers. To maximize entropy, we have

$$\frac{\partial \mathcal{L}_1}{\partial f_n} = 0,$$

which leads to

$$f_n = \omega_n e^{-\overline{\beta}_3 \epsilon_n}$$

where $\omega_n = e^{-\overline{\lambda}_3 - g_n^{-1}\left(\overline{\mu}_n + \sum_{n_2} \delta_*^{\Delta n_{12}} \overline{\mu}_{12}\right)}$, $\overline{\lambda}_3 = 1 + \lambda_3$, and $\overline{\mu}_n = -\sum_{n_1} \overline{\mu}_{12}$. The factor of $e^{-\overline{\beta}_3 \epsilon_n}$ is the counterpart of the equilibrium solution $e^{-\beta \epsilon_n}$, and $\omega_n$ reflects the nonequilibrium effect. Clearly, with $\omega_n$, $f_n$ is non-Boltzmannian. The associated maximization of entropy is more constrained than the chaotic case. Consequently, $S$ reaches the nonequilibrium maximum ($S_\mathrm{ne}$) that is less than the equilibrium maximum ($S_\mathrm{eq}$). If we take into consideration that the temperature field in Figure 6(a) tends to be heterogeneous (see Section 4.3), the derived $S_{\mathrm{ne}}$ would be even smaller.

**References**

1. I. Müller. *A History of Thermodynamics*, Springer, 2007.
2. Y. Kosmann-Schwarzbach. *The Noether Theorems*, Springer, 2010.
3. M. Kardar. *Statistical Physics of Particles*, Cambridge Univ. Press, 2007, pp. 16, 60, 74, 75, 99, 109.
4. U. Seifert. Stochastic thermodynamics, fluctuation theorems and molecular machines, *Rep. Prog. Phys.* **75** 126001 (2012)
5. H. S. Leff, A. F. Rex. *Maxwell's Demon: Entropy, Information, Computing*, Princeton Univ. Press, 1990
6. D. V. Averin, M. Möttönen, J. P. Pekola. Maxwell's demon based on a single-electron pump. *Phys. Rev. B* **84**, 245448 (2011).
7. R. P. Feynman, R. B. Leighton, M. Sands. *The Feynman Lecture Notes on Physics*, Basic Books, 2011, Vol. 1, Chapt. 46
8. L. Szilárd. On entropy reduction in a thermodynamic system by interference by intelligent subjects. *Zeitschrift fur Physik* **53**, 840–856 (1929) (translated in NASA Report No. TT-F-16723)
9. P. A. Skordos, W. H. Zurek. Maxwell's demon, rectifiers, and the second law: Computer simulation of Smoluchowski's trapdoor. *Am. J. Phys.* **60**, 876 (1992).
10. J. V. Koski, A. Kutvonen, T. Ala-Nissila, J. P. Pekola. On-chip Maxwell's demon as an information-powered refrigerator. *Phys. Rev. Lett.* **115**, 260602 (2015).
11. T. Sagawa, M. Ueda, Second law of thermodynamics with discrete quantum feedback control. *Phys. Rev. Lett*. **100**, 080403 (2008)
12. F. J. Cao, M. Feito, Thermodynamics of feedback-controlled systems. *Phys. Rev. E* **79**, 041118 (2009)
13. R. Landauer. Information is inevitably physical, in *Feynman and Computation*, Perseus Books, 1998, pp77-92
14. D. Mandal, C. Jarzynski. Work and information processing in a solvable model of Maxwell's demon. *Proc. Natl. Acad. Sci. U.S.A*. **109**, 11641 (2012)

15. Y. Qiao, Z. Shang. Producing useful work in a cycle by absorbing heat from a single thermal reservoir: An investigation on a locally nonchaotic energy barrier. *Physica A* **596**, 127105 (2022)
16. Y. Qiao, Z. Shang, R. Kou. Molecular-sized outward-swinging gate: Experiment and theoretical analysis of a locally nonchaotic barrier. *Phys. Rev. E* **104**, 064133 (2021)
17. The computer programs used in the current investigation are available at *http://mmrl.ucsd.edu/Z_Upload/Papers/SupplMater_StepRamp.zip*
18. G. Lente, K. Ősz. Barometric formulas: various derivations and comparisons to environmentally relevant observations. *ChemTexts* **6**, 13 (2020)
19. S. R. De Groot, P. Mazur. *Non-Equilibrium Thermodynamics*. Courier Corporation, 2013
20. Nicolas G. Hadjiconstantinou. The limits of Navier-Stokes theory and kinetic extensions for describing small-scale gaseous hydrodynamics. *Physics of Fluids* **18**, 111301 (2006).
21. James Jeans. *An Introduction to the Kinetic Theory of Gases*. Cambridge Univ. Press, 1982
22. M. N. Kogan. *Rarefied Gas Dynamics*. Springer, 1969.
23. G. Lebon, D. Jou. *Understanding Nonequilibrium Thermodynamics*, Springer, 2008
24. J. R. Dorfman. *An Introduction to Chaos in Nonequilibrium Statistical Mechanics*, Cambridge University Press, 1999.
25. A. Argun, A. Moradi, E. Pince, G. B. Bagci, A. Imparato, G. Volpe. Non-Boltzmann stationary distributions and nonequilibrium relations in active baths. *Phys. Rev. E* **94**, 062150 (2016).
26. W. Pauli, Thermodynamics and the kinetic theory of gases, in C. P. Enz (Ed.), *Pauli Lectures on Physics Vol. 3,* MIT Press, 1973.
27. J. C. Maxwell. Kinetic theory of gases. *Nature* **8**, 85 (1873)
28. J. C. Maxwell. *Theory of Heat*, Longmans, Green, and Co., London, 1872, p. 300
29. J. C. Maxwell. On the dynamical theory of gases. In: W. D. Niven (Ed.), *The Scientific Papers of James Clerk Maxwell Vol. II* (C. J. Clay & Sons, 1890) p. 76
30. Y. Qiao, Z. Shang. Spontaneous cold-to-hot heat transfer in Knudsen gas, under review. *https://doi.org/10.48550/arXiv.2312.09161*
31. N. D. Hari Dass. *The Principles of Thermodynamics*. CRC Press, 2013.

32. C. M. A. Brett, A. M. O. Brett. *Electrochemistry: Principles, Methods, and Applications*. Oxford University Press, 1993
33. H. Motschmann. Electrolytes at the air-water interface. In: G. Kreysa, K. Ota, R. F. Savinell (eds) *Encyclopedia of Applied Electrochemistry*. Springer, 2014.
34. K. Huang. *Statistical Mechanics*, John Wiley & Sons, 1987
35. P. Gaspard. *Chaos, scattering, and statistical mechanics*, Cambridge University Press, 1998
36. E. G. D Cohen, L. Rondoni. Particles, maps and irreversible thermodynamics. *Physica A* **306**, 117 (2002)
37. P. Gaspard, G. Nicolis and J.R. Dorfman, Diffusive Lorentz gases and multibaker maps are compatible with irreversible thermodynamics. *Physica A* **323**, 294 (2003)
38. H. Risken. *The Fokker-Planck Equation: Methods of Solution and Applications*, Springer-Verlag, 1984